\documentclass[final,3p,times,twocolumn,authoryear]{elsarticle}

\usepackage{hyperref}
\usepackage{amssymb}

\usepackage{xspace}
\usepackage{aas_macros}
\usepackage{xcolor}
\usepackage{hyperref}
\usepackage{rotating}
\usepackage{longtable}

\usepackage{array}
\newcolumntype{L}[1]{>{\raggedright\let\newline\\\arraybackslash\hspace{0pt}}m{#1}}
\newcolumntype{C}[1]{>{\centering\let\newline\\\arraybackslash\hspace{0pt}}m{#1}}
\newcolumntype{R}[1]{>{\raggedleft\let\newline\\\arraybackslash\hspace{0pt}}m{#1}}
\graphicspath{{images/}}

\journal{New Astronomy}

\usepackage{natbib} 
\usepackage{epsfig}
\usepackage{graphicx}
\usepackage{txfonts}
\usepackage{multicol}

\usepackage{amsmath}
 
\def\eflux{{\rm erg~cm$^{-2}$~s$^{-1}$}}
\def\nustar{\em NuSTAR}

\def\inte{{\em INTEGRAL}}
\def\xmm{{\em XMM-Newton}}

\def\chan{{\em Chandra}}

\def\rxte{{\em RXTE}}
\def\swift{{\em Swift}}

\def \inte {{\em INTEGRAL}}
\def \nicer {{\em NICER}}
\def \xmm {{\em XMM-Newton}}
\def \chandra {{$Chandra$}}

\def \ergsec{\hbox{erg s$^{-1}$}}

\def \hcm {\hbox {\ifmmode $ atom cm$^{-2}\else atom cm$^{-2}$\fi}}

\def\apjl{ApJL, }
\def\apjs{ApJS, }
\def \aap {A\&A, }

\journal{Journal of \LaTeX\ Templates}
\def\ltsima{$\; \buildrel < \over \sim \;$}
\def\simlt{\lower.5ex\hbox{\ltsima}}
\def\gtsima{$\; \buildrel > \over \sim \;$}
\def\simgt{\lower.5ex\hbox{\gtsima}}

\def\int {\emph{INTEGRAL}}
\def\sgr {SGR 1806$-$20}
\def\qui {1E 1547.0$-$5408}

\begin{document}

\begin{frontmatter}

\title{The {\inte} view of the pulsating hard X-ray sky: from accreting and transitional millisecond pulsars to rotation-powered pulsars and magnetars}


\author[address1]{A.~Papitto}
\ead{alessandro.papitto@inaf.it}

\author[address2]{M.~Falanga}

\author[address3,address4]{W.~Hermsen}

\author[address5]{S.~Mereghetti}

\author[address3]{L.~Kuiper}

\author[address7,address8,address9]{J.~Poutanen}

\author[address10]{E.~Bozzo}

\author[address11]{F.~Ambrosino}

\author[address13]{F.~Coti Zelati}

\author[address14]{V.~De Falco}

\author[address15]{D.~de Martino}

\author[address16]{T.~Di Salvo}

\author[address17,address5]{P.~Esposito}

\author[address10]{C.~Ferrigno}

\author[address18]{M.~Forot}

\author[address18]{D.~G{\"o}tz}

\author[address20]{C.~Gouiffes}

\author[address16]{R.~Iaria}

\author[address18]{P.~Laurent}

\author[address21]{J.~Li}

\author[address22]{Z.~Li}

\author[address24]{T.~Mineo}

\author[address25]{P.~Moran}

\author[address26,address19]{A.~Neronov}

\author[address5]{A.~Paizis}

\author[address13]{N.~Rea}

\author[address12]{A.~Riggio}

\author[address12]{A.~Sanna}

\author[address10]{V.~Savchenko}

\author[address27]{A.~S{\l}owikowska}

\author[address26]{A.~Shearer}

\author[address17,address5]{A.~Tiengo}

\author[address13,address28,address29]{D.~F.~Torres}

\address[address1]{INAF-Osservatorio Astronomico di Roma, via Frascati 33, Monte Porzio Catone, I-00078 Italy}
\address[address2]{International Space Science Institute (ISSI), Hallerstrasse 6, CH-3012 Bern, Switzerland}
\address[address3]{SRON Netherlands Institute for Space Research, Sorbonnelaan 2, NL-3584 CA Utrecht, the Netherlands}
\address[address4]{Anton Pannekoek Institute for Astronomy, University of Amsterdam, Science Park 904, NL-1098 XH Amsterdam, the Netherlands}
\address[address5]{INAF Istituto di Astrofisica Spaziale e Fisica Cosmica, Via A. Corti 12, I-20133 Milano, Italy}
\address[address7]{Tuorla Observatory, Department of Physics and Astronomy, 20014, University of Turku, Finland}
\address[address8]{Space Research Institute of the Russian Academy of Sciences, Profsoyuznaya str. 84/32, 117997, Moscow, Russia}
\address[address9]{Nordita, KTH Royal Institute of Technology and Stockholm University, Roslagstullsbacken 23, 10691, Stockholm, Sweden}
\address[address10]{ISDC, Department of Astronomy, University of Geneva, Chemin d'Ecogia 16, 1290, Versoix, Switzerland}
\address[address11]{INAF Istituto di Astrofisica e Planetologia Spaziali (IAPS), Via del Fosso del Cavaliere 100, I-00133 Rome, Italy}
\address[address13]{Institute of Space Sciences (ICE, CSIC), Campus UAB, Carrer de Can Magrans, E-08193, Barcelona, Spain}
\address[address14]{Research Centre for Computational Physics and Data Processing, Faculty of Philosophy \& Science, Silesian University in Opava, Bezručovo namesti 13, CZ-746 01 Opava, Czech Republic}
\address[address15]{INAF Osservatorio Astronomico di Capodimonte, Salita Moiariello 16, I-80131 Napoli, Italy}
\address[address16]{Universit\`a degli Studi di Palermo, Dipartimento di Fisica e Chimica - Emilio Segré, via Archirafi 36, I-90123 Palermo, Italy}
\address[address17]{Scuola Universitaria Superiore IUSS Pavia, piazza della Vittoria 15, 27100, Pavia, Italy}
\address[address18]{CEA, IRFU, Service d'Astrophysique, Orme des Merisiers, 91191 Gif-sur-Yvette, France}
\address[address20]{Laboratoire AIM, UMR 7158 (CEA/Irfu, CNRS/INSU, Université Paris VII), CEA Saclay, Bt. 709, F-91191 Gif-sur-Yvette Cedex, France}
\address[address21]{Deutsches Elektronen Synchrotron DESY, D-15738 Zeuthen, Germany}
\address[address22]{Department of Physics, Xiangtan University, Xiangtan, 411105, PR China}
\address[address24]{INAF Istituto di Astrofisica Spaziale e Fisica Cosmica Palermo, Via U. La Malfa 153, I-90146 Palermo, Italy}
\address[address25]{Centre for Astronomy, School of Physics, National University of Ireland Galway, University Road, Galway, Ireland}
\address[address26]{Astronomy Department, University of Geneva, Ch. d'Ecogia 16, 1290, Versoix, Switzerland}
\address[address12]{Dipartimento di Fisica, Universit\`a di Cagliari, SP Monserrato-Sestu, Km 0.7, I-09042 Monserrato, Italy}
\address[address19]{APC, Astroparticule et Cosmologie CNRS/IN2P3, CEA/IRFU, 10 rue Alice Domon et Leonie Duquet, F-75013, Paris, France}
\address[address27]{Institute of Astronomy, Faculty of Physics, Astronomy and Informatics, 
Nicolaus Copernicus University in Toru\'n, Grudziadzka 5, PL-87-100 Toru\'n, Poland}
\address[address28]{Institut d'Estudis Espacials de Catalunya (IEEC), E-08034 Barcelona, Spain}
\address[address29]{Instituci\'o Catalana de Recerca i Estudis Avançats (ICREA), E-08010 Barcelona, Spain}

\begin{abstract}

In the last 25 years,  a new generation of X-ray satellites  imparted a significant leap forward in our knowledge of X-ray pulsars. The discovery of accreting and transitional millisecond pulsars proved that disk accretion can spin up a neutron star to a very high rotation speed. The detection of MeV-GeV pulsed emission from a few hundreds of rotation-powered pulsars probed particle acceleration in the outer magnetosphere, or even beyond. Also, a population of two dozens of magnetars has  emerged. {\inte} played a central role to achieve these results by providing instruments with high temporal resolution up to the hard X-ray/soft $\gamma$-ray band and a large field of view imager with  good angular resolution to spot hard X-ray transients. In this article, we review the main contributions by {\inte} to our understanding of the pulsating hard X-ray sky, such as the discovery and characterization of several accreting and transitional millisecond pulsars, the generation of the first catalog of hard X-ray/soft $\gamma$-ray rotation-powered pulsars, the detection of polarization in the hard X-ray emission from the Crab pulsar, and the discovery of persistent hard X-ray emission from several magnetars.
\end{abstract}

\begin{keyword}
\texttt{accretion disks}\sep  \texttt{magnetars}\sep \texttt{neutron stars}\sep \texttt{pulsars}\sep \texttt{X-rays: binaries}\sep \texttt{X-rays: bursts}
\end{keyword}

\end{frontmatter}


\section{Introduction}

Since 1967, the pulsar phenomenon has provided the largest share of
information we have obtained so far on the properties of neutron stars
\citep[henceforth NS;][]{hew68,pac67,gol68}. By now, emission
coherently modulated by the NS rotation has been observed at all
wavelengths, from the radio to the very high energy domains. A few
physical mechanisms can produce pulsations observed at X-ray and soft
gamma-ray energies covered by {\inte}; magnetically channeled
accretion of matter transferred from a companion star in a binary
system, spreading of the thermonuclear burning front over the NS
surface resulting from the ignition of accreted matter (so-called
type-I X-ray burst oscillations), pulsed emission powered by the
rotation of the strong electro-magnetic field anchored to the NS
surface and/or the dissipation of such a field in magnetars. Observing
pulsars has been crucial to measure fundamental properties of NSs
(such as masses, radii, magnetic fields) and understand their
evolution (see, e.g., \citealt{gho07}).

In this chapter we review the main results achieved by the {\inte}
mission \citep{win03} on accreting and transitional millisecond
pulsars, rotation-powered pulsars and magnetars. In this regard, the
hard X-ray (20~keV$-$1~MeV) imager IBIS/ISGRI \citep{ube03,leb03}
played a major role, as it combined a large field of view
($29^{\circ}\times29^{\circ}$ with a fully coded field of
$8^{\circ}\times8^{\circ}$), fine angular resolution ($12'$ full-width
half-maximum) and high temporal resolution \citep[$60\,\mu$s;][]{kui03}. The X-ray (3$-$35~keV) monitor JEM-X \citep{lun03}
complemented these properties providing a better angular resolution of
$3'$ and sensitivity to softer energies, although with a smaller field
of view (the fully illuminated part is $\sim4.8^{\circ}\times
4.8^{\circ}$-wide).

The structure of the article is the following. In Sect.~\ref{sec:amsp}
we review the {\inte} contribution to the study of accreting
millisecond pulsars, Gyr-old and relatively weakly magnetized ($\simeq
10^{8}-10^{9}$~G) NSs that were spun-up to their current fast spin
period ($P\simeq 1-10$~ms) by the accretion of matter transferred from
a low mass companion star ($\le M_{\odot}$). These sources also show
bursts of soft X-rays caused by the thermonuclear burning of the
material accreted on the surface and the {\inte} results are
summarized in Sect.~\ref{sec:amspburst}. We refer the reader to the
article by \citet{sazonov20} for the results obtained for slower and/or
non-pulsating NSs in low-mass X-ray binaries, and to the article by
\citet{kretschmar20} for the case of X-ray pulsars in high-mass X-ray
binaries. Sect.~\ref{sec:trans} describes the
role played by {\inte} in discovering and characterizing transitional
millisecond pulsars, a small sample of sources that are able to
alternate between phases of emission as a low-mass X-ray binary and
regimes characterized by radio pulsar emission. Studies of pulsars
powered by the rotation of their magnetic field, including both slower
($P\sim0.1-10$~s) and strongly magnetized ($B\sim10^{11}-10^{13}$~G)
{\it classical} pulsars and recycled millisecond pulsars
($P\sim\,1-10\,$ms; $B\sim10^8$~G) are presented in
Sect.~\ref{sec:radiopulsars}. Finally, {\inte} observations of NSs
powered by the dissipation of their intense ($B\sim10^{13}-10^{14}$~G)
magnetic field (so-called magnetars) are summarized in
Sect.~\ref{sec:magnetars}.

\section{Accreting millisecond pulsars}
\label{sec:amsp}
Accreting millisecond pulsars (AMSPs in the following) are NSs that
transiently accrete the plasma captured from a low-mass companion star
($M_2 \le M_{\odot}$) via Roche-lobe overflow (\citealt{wijnands1998};
see \citealt{patruno2012,campana2018,disalvo20} for reviews). Their
magnetic field ($B_p\simeq10^8-10^9$~G) is strong enough to truncate
the disk in-flow before the plasma reaches the surface of the NS. The
in-falling matter is then channeled by the magnetic field of the NS to
the magnetic polar regions of the NS surface. As long as the magnetic
and spin axes are misaligned and the emission beam crosses the line of
sight, this produces coherent pulsations mainly observed in the X-ray
domain.

The spin periods of AMSPs range between 1.6 and $\sim
10$~ms. According to the recycling evolutionary model
\citep{bisnovatyikogan1974,alpar1982,radhakrishnan1982}, such an
extremely quick rotation is achieved during a prolonged ($\approx
0.1-1$~Gyr) X-ray bright phase of accretion of matter lost by a
low-mass companion star.  The $\sim$300 millisecond radio pulsars
known to date are then assumed to be the descendants of low mass X-ray
binaries (LMXBs in the following). In these systems, a radio pulsar
turns on as soon as the pressure of the pulsar wind inhibits the
in-fall of matter lost by the companion and accretion ceases (see also
Sect.~3).

\begin{figure*}
\centering 
\includegraphics[width=10cm]{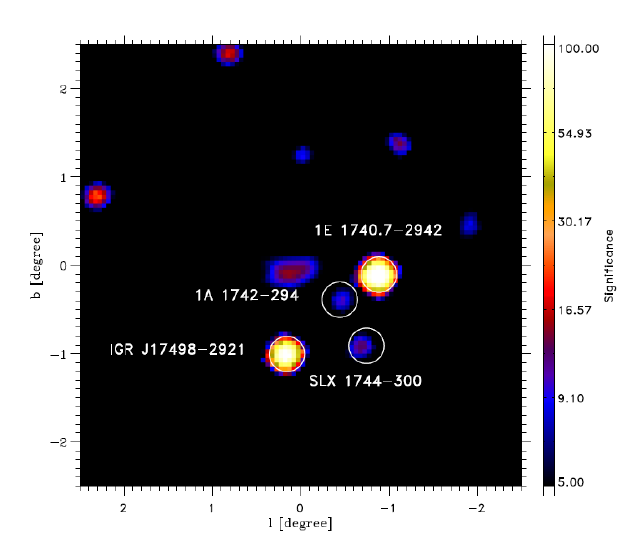}
\caption{IBIS/ISGRI 20-100 keV sky image of the field around IGR~J17498-2921
obtained during  observations performed at the time of the source discovery in 2011, giving an effective exposure of 210 ks. Credit: Falanga et al., A\&A, 545, A26 (2012), reproduced with permission (ESO).}
\label{fig:fig1}
\end{figure*}

So far, AMSPs have been found in binaries hosting either a main
sequence star ($M_2 \sim 0.1-0.5$ M$_{\odot}$, $P_{orb}\approx$ a few
hours), a brown dwarf ($M_2\approx 0.05$ M$_{\odot}$, $P_{orb}\simeq
1-2$ hr) or a white dwarf ($M_2\approx 0.01$ M$_{\odot}$,
$P_{orb}\simeq 40$ min). These binary characteristics are similar to
those of binary millisecond radio pulsars whose signal is irregularly
eclipsed by matter ejected by the pulsar wind and engulfing the binary
system. These eclipsing radio pulsars are dubbed either black widow
($M_2 \approx 0.05$ M$_{\odot}$; \citealt{fru88,dra19}) or redback
pulsars ($M_2 \simeq 1$ M$_{\odot}$: \citealt{dam01,str19}) depending on
the mass of the companion star, and are generally considered to share
a close evolutionary link with AMSPs \citep{klu88,vdh88,rud89}.

Discovering AMSPs and measuring their spin evolution is crucial to
understand what is the maximum spin that can be reached by a NS
through accretion. In turn, this indirectly probes whether continuous
gravitational-wave spin-down torques are required to limit the
accretion driven spin up to the minimum observed period of a
millisecond pulsar, $\approx 1.5$~ms
\citep{chakrabarty2008,papitto2014b,patruno2017}. The X-ray pulsed
emission of AMSPs is emitted close to the surface of a rapidly
rotating object that attains a speed of up to $\approx 15\%$ of the
speed of light at the equator. General and special relativity effects
shape the energy and trajectory of X-ray photons in a way that can be
disentangled through X-ray pulse profile fitting. This makes AMSPs
among the best candidates to measure simultaneously the mass and
radius of a NS and draw constraints on its equation of state
(\citealt{pou03}; see also \citealt{wat16} for a recent review).

\subsection{Discovery and follow-up of AMSPs with {\inte}}

\begin{table*}[h!]
  \small
\centering
\begin{tabular}{||l c c c l||} 
 \hline
  & $P_{spin}$ (ms) & $P_{orb}$ (hr) &  Outburst year & References \\ [0.5ex] 
 
 \multicolumn{5}{|| l ||}{{\inte} sources}\\ 
 \hline
 IGR~J00291+5934 & 1.7 & 2.46 & 2004, '15 & \citet{sha05,mf05b};\\
 & & & & \citet{falco17b} \\
 IGR~J17511$-$3057 & 4.1 & 3.47 &  2009 & \citet{mfalanga11} \\
 IGR~J17498$-$2921 & 2.5 & 3.84 & 2011 & \citet{mfalanga12} \\
 IGR~J17480$-$2446 & 90 & 21.3 & 2011 &  \citet{bor10,fer10b};\\
 & & & & \citet{str10, pap11}\\
 IGR~J18245$-$2452 & 3.9 & 11.0 & 2013 & \citet{papitto2013,fer14};\\
 & & & & \citet{def17} \\
 IGR~J17062$-$6143 & 6.1 & $> 0.3$ &  2008 &  \citet{strohmayer2017} \\
 IGR~J16597$-$3704 &  9.5 & 0.77 & 2017 & \citet{sanna18b} \\
 IGR~J17379$-$3747 & 2.1 & 1.88 & 2018 &  \citet{sanna18, strohmayer2018} \\
 IGR~J17591$-$2342 & 1.9 & 8.80 & 2018 &  \citet{sanna18c} \\ [1ex] 

  \multicolumn{5}{|| l ||}{Other sources}\\
  \hline
  SAX~J1808.4$-$3658 & 2.5 & 2.01 & 2008, '15, '19 & \citet{delsanto15,patruno2017b};\\
  & & & & \citet{ferrigno19} \\ 
  XTE~J1751$-$305 & 2.3 & 0.71 & 2005, '07, '09 & \citet{grebenev05,falangaatel07};\\
  & & & & \citet{chenevez09} \\
  XTE J1807$-$294 & 5.3 & 0.67 & 2003 & \citet{mf05}\\
  HETE~1900.1-2455 & 2.7 & 1.39 & 2005 & \citet{mf07}\\
  SAX J1748.9-2021 & 2.3 & 8.77 & 2015, '17 & \citet{kuulkers15,gesu17,li18}\\
  Swift\,J1749.4-2807 &  1.9 & 8.82 &  2010 & \citet{pavan10,chenevez10,ferrigno11}\\
  MAXI\,J0911-655 & 2.9 & 0.74 & 2016 & \citet{sanna17, bozzo16, ducci16} \\
   \hline
\end{tabular}
\caption{AMSPs observed by {\inte}.}
\label{table:amsp}
\end{table*}

The two dozens of AMSPs discovered so far are all X-ray transients. They undergo a few weeks-long X-ray outbursts and spend most of the time in quiescence, although episodes of X-ray activity lasting up to a few years have also been observed. Recurrence times range from a few months to more than 15 years, for sources which have been observed only once, so far. During outbursts, the mass accretion rate rarely exceeds a few per cent of the Eddington rate ($L_X\approx10^{36}-10^{37}$~erg~s$^{-1}$), whereas in quiescence the luminosity is much lower ($L_X\simlt10^{32}$~erg~s$^{-1}$). The discovery of new systems of this class requires instruments with a large field of view and a good sensitivity.

The instruments on-board {\inte} perfectly satisfy these
requirements. They managed to discover the X-ray outbursts of eight
sources that were later identified as AMSPs (out of a total of 22; see
Table \ref{table:amsp}, \citealt{fal13}, and references therein), and
to study their X-ray emission from the onset of the outbursts nearly
down to the return in quiescence. Since the launch of {\inte} the
large field of view of the IBIS/ISGRI imager has been exploited to
monitor regularly the Galactic bulge \citep{a436,kuulkers07}, where
most of the transient LMXBs are expected. The highly eccentric long
orbit and the special pointing strategy adopted by \inte\ allowed
IBIS/ISGRI to accumulate several thousands of kiloseconds of
observations in each monitored region, with a few-hours long
uninterrupted coverage, and achieve a hard ($20-200$~keV) X-ray flux
sensitivity as low as a few $\times$~10$^{-11}$~\eflux even in crowded
regions. For a distance between 5-8~kpc, this limiting sensitivity
corresponds to a luminosity of $\approx 10^{35}$~\ergsec, which is
usually reached by AMSPs already during the earliest stages of their
outbursts, and attained again towards the end of the outbursts before
the switch back to quiescence. IBIS/ISGRI observations were also often
complemented by data in the soft band covered by JEM-X at similar
sensitivity.

\begin{figure}[t!]
\centering 
\includegraphics[width=\columnwidth]{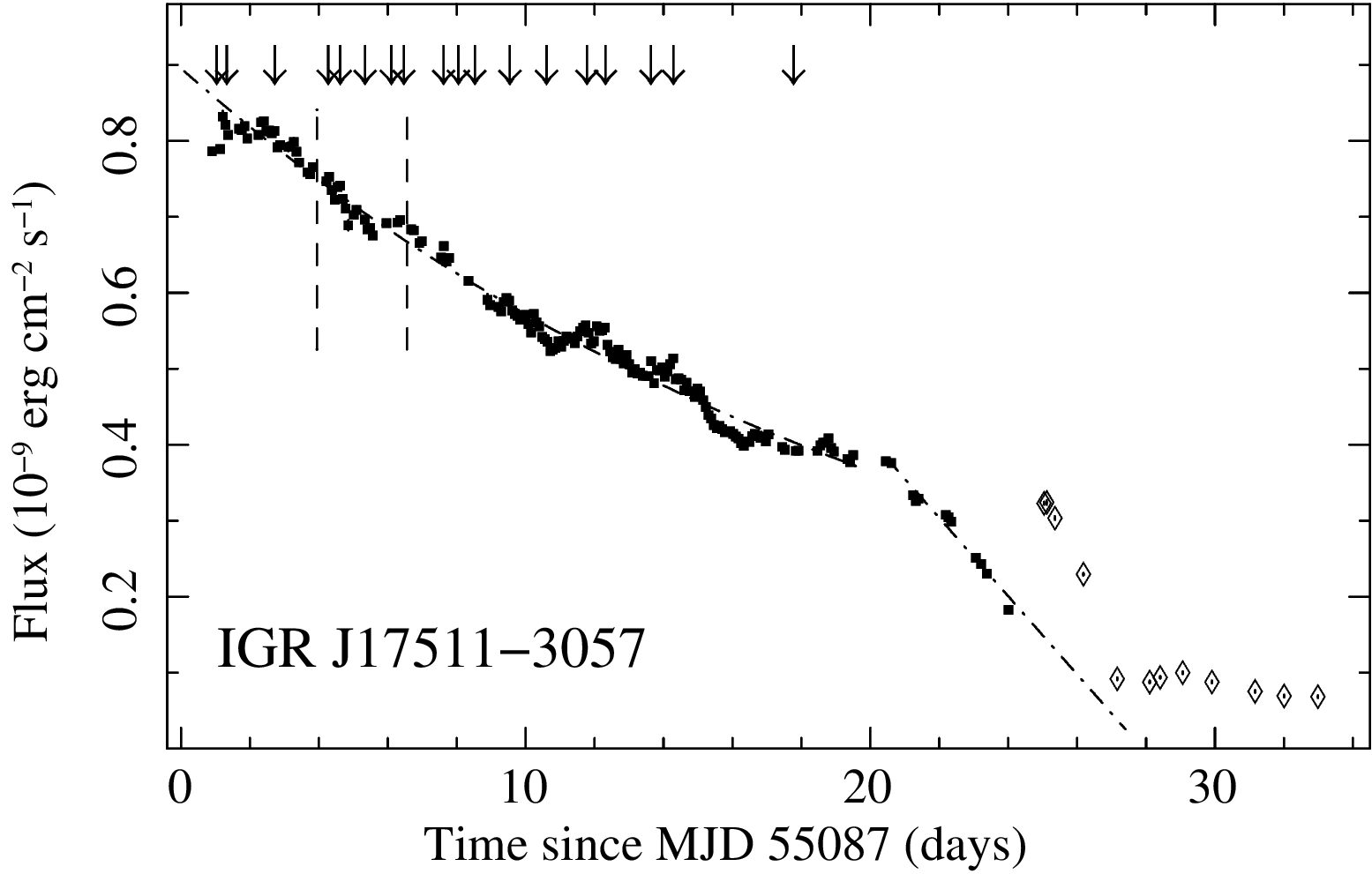}
\caption{The light curve of the outburst of IGR\,J17511-3057 observed in 2009  \citep{mfalanga11}. The typical exponential decay profile is clearly visible, with a break occurring about 20 days after the onset of the event. The light curve is obtained from the {\rxte}/PCA (2$-$20 keV) data, with the vertical dashed lines indicating the interval of the \inte\ observations. The arrows mark the times of the detected X-ray bursts. The diamonds refer to the observations in which both IGR J17511$-$3057 and XTE J1751$-$305  were active and the instruments on-board
{\it RXTE} were unable to separate the contribution of the two sources. Credit:
Falanga et al., A\&A, 529, A68 (2011), reproduced with permission (ESO).}
\label{fig:inte_lc}
\end{figure}

These features, complemented by the good angular resolution of the two
instruments, proved crucial to discover AMSPs and disentangle their
emission from foreground sources (see Fig.~\ref{fig:fig1}). These
imaging capabilities were particularly important to identify
IGR~J00291+5934, which is located only $\sim18'$ from the close-by
persistent intermediate polar V609~Cas \citep{mf05b}, and
IGR~J17511-3057, which is located only $\sim20'$ away from the other
AMSP XTE~J1751-305 and whose discovery outburst was partly
contaminated by a faint activity episode of the latter source
(\citealt{mfalanga11}, see Fig.~\ref{fig:inte_lc}). The good angular
resolution of IBIS/ISGRI and JEM-X has also proven fundamental to
clearly distinguish the type-I X-ray bursts emitted by the AMSP
IGR~J17498-2921 from those going off in the nearby bursters
SLX~1744-300/299 and 1A~1742-294, located at $0.9^{\circ}$ and
$0.86^{\circ}$, respectively, from the AMSP \citep{mfalanga12}. In
2010 {\inte} also discovered the LMXB IGR~J17480-2446 in the globular
cluster Terzan 5 \citep{bor10,fer10b}. The source was subsequently
identified as a 90~ms pulsar \citep{str10} orbiting a low-mass
companion star in a 21~hr orbit (\citealt{pap11}, see
Sect.~\ref{sec:amspburst} for details).

\inte\ observations have also been extensively carried out to follow up AMSPs first detected by other facilities and which underwent one or multiple outbursts since the beginning of the mission science operations in 2002.  In a few cases (e.g. XTE~J1751$-$305, SAX J1748.9-2021, SAX J1808.4-3658), the \inte\ seasonal pointings toward the Galactic bulge detected the onset of some of the outbursts of AMSPs already discovered. 
The LMXB nature of Swift\,J1749.4-2807 was first established during its 2010 outburst announced by \inte\ \citep{pavan10,chenevez10}; the source was later indentified as a 1.9~ms AMSP which showed 1.7~ks-long X-ray eclipses \citep[see, e.g.,][]{ferrigno11, altamirano11}, making it the first and only eclipsing AMSP discovered so far, with important consequences for the determination of the NS mass and radius \citep{jonker13}. The first outburst of  MAXI\,J0911-655, was also observed by \inte\ \citep{sanna17, bozzo16, ducci16}, which also provided a long term monitoring of the source reporting on the discovery of significant hard X-ray emission more than 450~days after the onset of the event \citep{victor17}.

\subsection{The X-ray light curves}
\label{sec:amsplc}

IBIS/ISGRI and JEM-X provided long term monitoring data with high
cadence which have been important to extract light-curves of AMSP
outbursts and identify the physical mechanisms driving the outburst
onset and decay. The observed light curves were generally
characterized by a fast rise (a few days) and an exponential decay (up
to several weeks) which terminated with a break followed by a linear
decay extending down to the limiting observable flux for both
IBIS/ISGRI and JEM-X (see Fig.~\ref{fig:inte_lc}).  This behaviour has
been commonly interpreted in terms of the disk instability model, in
which the irradiation of the accretion disk by the central X-ray
source plays a key role in the shaping of the light curve profile
\citep{king98}. Modelling of the light curves observed by the {\it
  Rossi X-ray Timing Explorer} ({\rxte}, \citealt{rxte}) showed that the
timescale of the decay and its luminosity at a characteristic time are
linked to the outer radius of the accretion disk \citep{pow07}.  The
break observed during the X-ray flux decay at the end of the outburst
is thought to be associated with the lowest X-ray luminosity at which
the outer disk region can be kept in a hot high-viscosity state by the
centrally illuminating source. When the outer disk region enters the
cool low-viscosity state, the mass accretion rate onto the compact
object is effectively cut-off and the source starts its return to
quiescence. The application of this model to the {\inte} data gave
compatible results \citep{mf05b, mfalanga11, mfalanga12, ferrigno11}.

So far, only IGR\,J00291+5934 has shown a double-peaked outburst \citep{lewis10, hartman11}, while a few other AMSPs have undergone ``re-flares'' toward the later stages of the return to quiescence. The mechanism(s) driving these re-brightening episodes is still a matter of debate \citep[see, e.g.,][and references therein]{patruno16,bult19}. The typical luminosity at which the re-flares occur is close to the detection limit for both IBIS/ISGRI and JEM-X and thus these events are hardly observable by {\inte}.

\subsection{The hard X-ray spectra of AMSPs}
\label{sec:hardspecamsp}
\begin{figure}
\centering 
\includegraphics[width=\columnwidth]{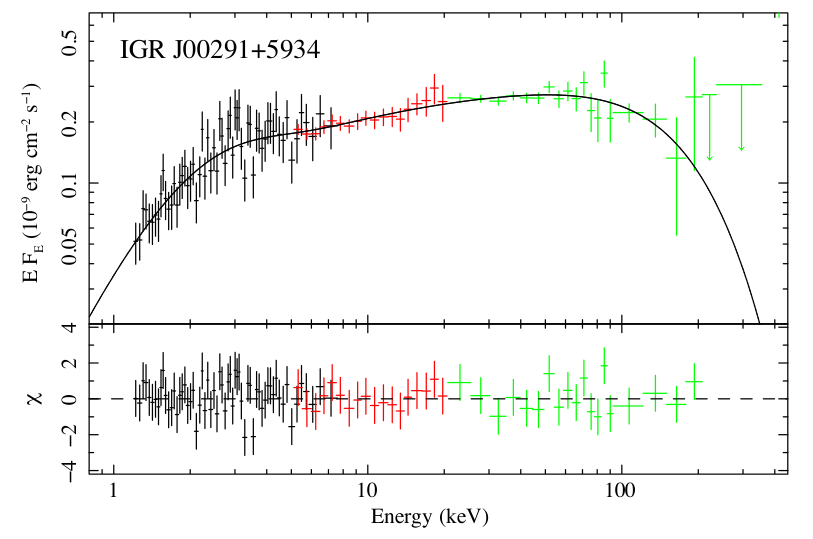}
\caption{The unfolded broad-band spectrum of the AMSP IGR J00291+5934 measured by JEM-X (red points) and IBIS/ISGRI (green points) during an outburst exhibited in 2015. {\swift}/XRT data were also used for the analysis and are shown as black points. The best fit is obtained with the {\sc compps} model (solid black line), resulting in a plasma temperature of $kT\sim50$~keV. The bottom panel shows the residuals from the best fit. Credit: De Falco et al., A\&A, 599, A88
(2017), reproduced with permission (ESO).}
\label{fig:igr_spe}
\end{figure}

The spectra of AMSPs in outburst measured by {\inte}  are typically hard and dominated by a power law $dN/dE\propto E^{-\Gamma}$, with photon index $\Gamma\sim2$, extending up to 100 keV or beyond \citep{pou06}. Most likely, these hard X-ray photons originate from the accretion columns above the polar caps, where electrons energized by the shock between the in-falling plasma and  the NS surface  up-scatter the surface soft photons to higher energies \citep{gierlinski2002,gierlinski2005}.  In most cases, the combined IBIS/ISGRI+JEM-X spectra could be well fit with a thermal Comptonization model {\sc compps} \citep{ps96} in the slab geometry assumed for the accretion columns \citep{mf05b,mfalanga11,mfalanga12,mf05,mf07,ferrigno11}. The main model parameters are the Thomson optical depth $\tau_{\rm T}\sim 1-3$ across the slab, the electron temperature $kT_{\rm e}\sim 25-50$ keV,
the temperature $kT_{\rm bb}\sim 0.3-1.0$ keV of the soft-seed blackbody photons (assumed to be injected from the bottom of the slab), and the emission area $A_{\rm bb}\sim 20$ km$^2$. Fig.~\ref{fig:igr_spe} shows a typical  broad-band AMSP spectrum observed during the decay of an outburst, fitted with a {\sc compps} model. The spectrum observed during an outburst of the transitional millisecond pulsars IGR J18245$-$2452 is instead significantly harder, with a photon index $\Gamma\sim1.3$ and seed photons coming from a larger region and characterized by a  lower temperature (\citep{def17}, see Sect.~\ref{sec:tmspj18245}). {\inte} observations of Aql X-1, which has been detected as an AMSP only for a couple of minutes \citep{cas08}, also caught the source in a state characterized by a very hard spectrum extending up to 150~keV \citep{rod06}.

\begin{figure*}[h!]
\centering 
\includegraphics[width=13.0cm]{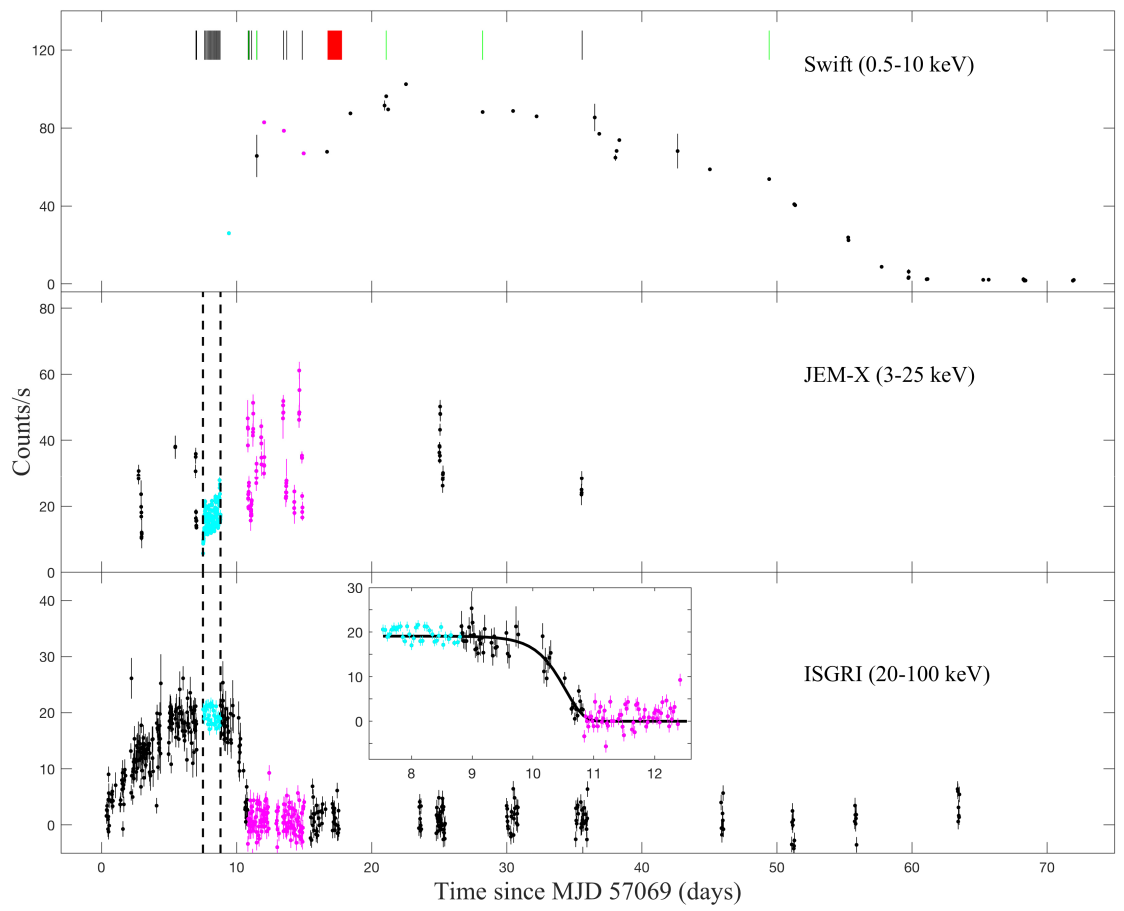}
\caption{The light curve from the 2015 outburst of the AMSP SAX\,J1748.9-2021. The top panel shows data collected with {\swift}/XRT, while the two panels below report the JEM-X and IBIS/ISGRI data. The spectral transition from hard (light blue) to soft (magneta) state is clearly visible in the \inte\ data (see the inset of the bottom panel). The figure also shows the time of a dedicated 100~ks-long \inte\ ToO observation (marked with vertical dashed lines) and all the thermonuclear bursts recorded during the outburst with \inte,\ \swift,\ and \xmm\ (black, green, and red vertical lines, respectively). Credit: Li et al., A\&A, 620,
A114 (2018), reproduced with permission (ESO). }
\label{fig:inte_lc2}
\end{figure*}

The spectra of several {\inte}-detected AMSPs have been also observed by using combined {\inte} and (quasi-)simultaneous {\xmm} \citep{xmm} and {\nustar} \citep{nustar} observations. Although they have been clearly found to be dominated by hard power-law components, typically ascribed to the Comptonization of seed thermal photons in agreement with previous \inte\ results, the extension into the soft X-ray regime frequently revealed the presence of additional thermal components originating from hot spots on the NS surface or from the inner disk boundary. In addition, a broad iron line produced by the reflection of the X-ray photons onto the inner regions of the accretion disk surrounding the compact object was sometimes also observed \citep{pap09, cac09, sanna17b, sanna18c, dis19}. In at least one case, high spectral resolution observations carried out with the gratings on-board \chan\ were able also to detect outflows from the outer regions of the accretion disk  \citep{nowak19}, and to provide hints of an expanding hot corona with high outflow velocities also in the case of IGR J00291+0034 \citep{pai05}.

Little spectral variability, if any, has been generally observed
during the course of an outburst. SAX J1748.9-2021 in the globular
cluster NGC 6440 clearly made an exception. It has been observed to
switch between hard and soft spectra \citep{pat09} such as occurs in
the so-called ``Atoll'' sources \citep{has89}. The combined extensive
monitoring performed with IBIS/ISGRI and JEM-X during the source
outburst in 2015 was able to efficiently catch one of these spectral
state changes (\citealt{li18}; see Fig.~\ref{fig:inte_lc2}); a dramatic
transition from a hard to a soft state state occurred over roughly
half a day, about ten days after the onset of the outburst. In that
outburst the source reached a peak luminosity of about $5\times
10^{37}$ erg~s$^{-1}$, a value higher than usually observed from
AMSPs. After the state change, the spectrum observed with {\xmm}
appeared to be very soft with a Comptonization electron temperature
around 2 keV \citep{pin16}. The (quasi) simultaneous spectrum observed
with {\inte} also revealed the presence of a hard power-law component
with a photon index of 2.3.  Similar hard tails are often observed in
Z-sources (see e.g. the {\inte} spectrum of Sco X-1;
\citealt{dis06,rev14}) and atoll sources in the soft state (see e.g. the
IBIS/ISGRI spectrum of GX 13+1 \citealt{pai06}, and the {\xmm}/{\inte}
spectrum of GX 3+1; \citealt{pin15}) and are interpreted in terms of
Comptonization of photons off electrons with a non-thermal
distribution of velocity. This non-thermal component may be related to
high energy electrons injected in the Comptonization region by a
(failed?) jet or powered by magnetic reconnections close to the
accretion disk. During the outburst occurred in October 2017, SAX
J1748.9-2021 showed instead a {\it standard} outburst X-ray luminosity
of $3\times 10^{36}$ erg~s$^{-1}$ and a more common hard spectrum
(photon index $\Gamma\sim1.6-1.7$, and electron temperature of 20 keV,
\citep{pintore18}). This demonstrates the importance of a high-energy
monitor such as {\inte} for addressing the spectral state of these
sources and individuate peculiar ones.

\begin{figure*}[t!]
\centering 
\includegraphics[width=13cm]{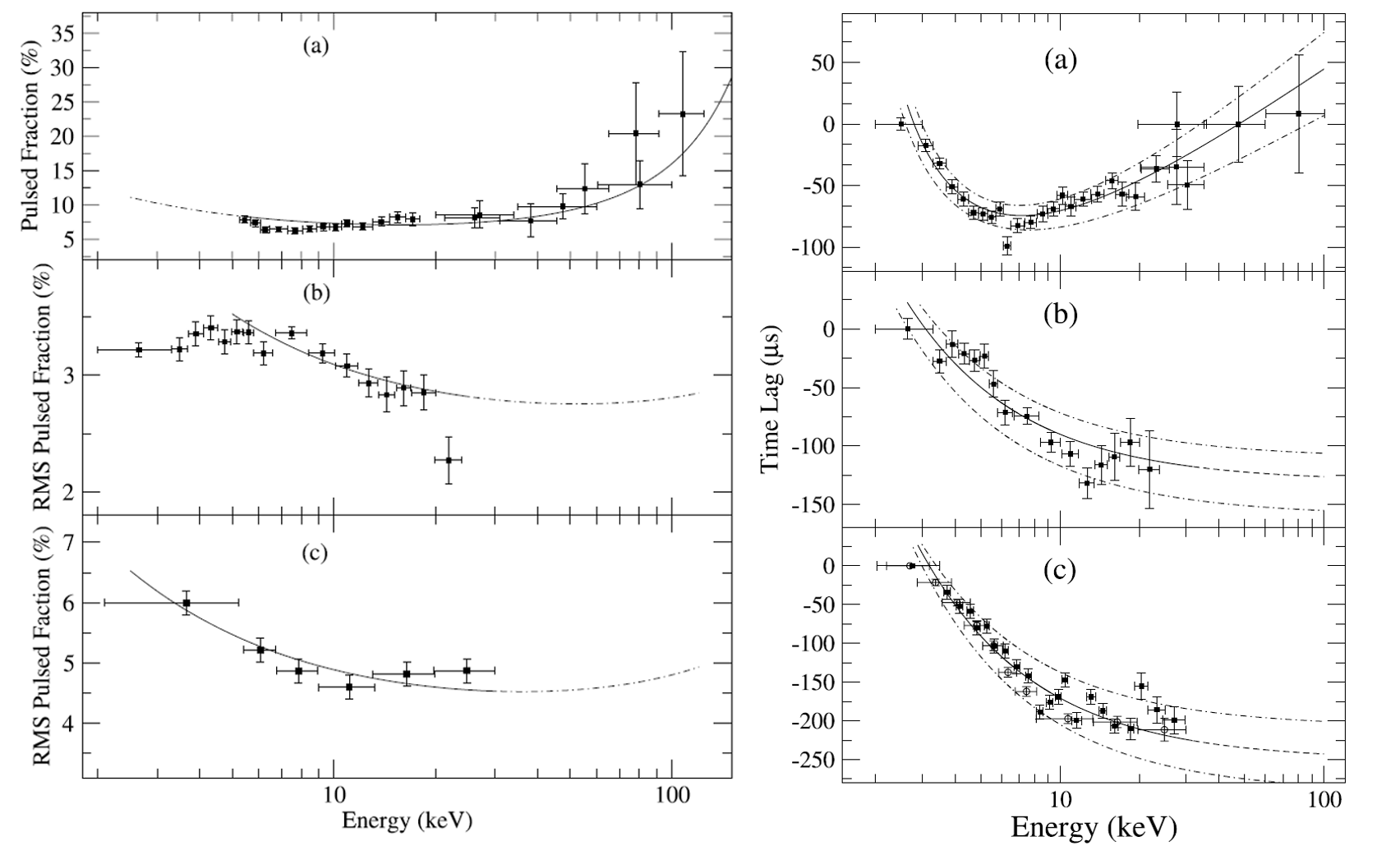}
\caption{The pulsed fraction (panel (a), left) and the time lags of the hard pulse (panel (a), right) measured during the 2005 outburst of the AMSP IGR J00291+5934 using data from the {\rxte}/PCA, {\rxte}/HEXTE, {\inte}/JEM-X, and {\inte} IBIS/ISGRI. Panel (b) and (c) show the same quantities for SAX J1808.4$-$3658 and XTE~J1751$-$305, as measured by {\rxte}. (Credit: Falanga \& Titarchuk, ApJ, 661, 2 (2007), reproduced with permission of the AAS).} 
\label{fig:j00}
\end{figure*}

 \subsection{The hard X-ray pulse profiles}
\label{sec:amsppulse}
AMSPs are relatively faint X-ray transients and the amplitude of their pulsations is $\simlt 10$ per cent, \citep{patruno2012}. For this reason, usually X-ray pulsations could not be discovered independently by {\inte} without a previous detection by a large area focusing or collimated instruments, as those on-board {\rxte}, {\xmm}, and more recently {\nustar} and {\it The Neutron Star Interior Composition Explorer Mission} ({\nicer}, \citep{nicer}). 
Only during the 2015 outburst of IGR J00291+5934, did \inte\ data
statistics reach a sufficient level to measure the pulsar ephemeris
directly from the IBIS/ISGRI event files \citep{kuiper2015b},
obtaining parameters compatible with those derived by {\rxte}.  Once
the spin and orbital parameters of an AMSP were provided, the {\inte}
data could be folded to increase the signal-to-noise ratio and carry
out a pulse timing analysis in a very broad energy range, covering up
to a few hundreds of keV.  In this way, IBIS/ISGRI detected X-ray
pulsations up to 150~keV from IGR~J00291+5934 \citep{mf05b},
IGR~J17511-3057 \citep{mfalanga11}, IGR~J17498-2921
\citep{mfalanga12}, IGR\,J18245-2452 \citep{def17}, and
IGR\,J17062-6143 \citep{kuiper20}, while an accurate analysis of the
data has not been published yet for recently discovered IGR AMSPs,
such as IGR J16597-3704, IGR J17379-3747 and IGR J17591-2342.  The
measurements obtained by IBIS/ISGRI permitted to perform an analysis
of the pulsed fractions in a hard ($> 20$~keV) X-ray domain previously
poorly covered by other instruments. Different trends were observed
from source to source. IGR~J17511-3057, IGR~J17498-2921 and
IGR\,J18245-2452 showed a constant, or slightly decreasing pulsed
fraction above 10~keV, compatible with the results obtained by {\rxte}
for SAX J1808.4-3658 and XTE J1751$-$305 (see left panels of
Fig.~\ref{fig:j00} labelled as (b) and (c); see \citealt{fal07} and
references therein). A decrease of the pulsed fraction at high
energies has been explained in the context of the two pulsed
components assumed to originate from the hot-spot on the NS surface
and from the accretion column above the poles (\citealt{gierlinski2002,
  gierlinski2005}, see Sect.~\ref{sec:hardspecamsp}). The soft
blackbody radiation from the hot-spots is more beamed along the axis
than the hard photons produced by Compton up-scattering in the
accretion columns, naturally accounting for the lower pulsed fraction
observed at higher energies \citep{pou03}. On the other hand, a clear
increase of the pulsed fraction with energy above $\sim50$~keV was
observed from IGR~J00291+5934 (see left panel labelled as (a) in
Fig.~\ref{fig:j00}). Embedding of the accretion columns in a Compton
cloud was proposed to explain this \citep{fal07}. In fact, the
electron cross section and the resulting Compton optical depth
decrease at high energies, allowing a larger fraction of the pulsed
harder X-rays to reach the observer un-scattered. Note that a trend of
increasing pulsed fraction with energy, albeit already at softer
X-rays, was observed by other missions (e.g., {\rxte}, {\xmm} and
{\nustar}) also from SAX J1748.9$-$2021 \citep{pat09,san16}, Swift
J1756.9$-$2508 \citep{pat10,san18} and, to a lesser extent, Swift
J1749.4$-$2807 \citep{ferrigno11}.

\inte\ crucially contributed also to reveal the complex dependence of
time lags of pulsed photons with energy (see right panel of
Fig.~\ref{fig:j00}, taken from \citealt{fal07}). The measured soft
lags mean that the low-energy pulses are delayed relative to pulses at
higher energies. Below $\sim 10$~keV, soft photons generally lag the
hard photons, although the opposite behavior was observed during an
outburst of the transitional millisecond pulsar IGR~J18245-2452 (see
Sect.~\ref{sec:tmspj18245}). The observed soft phase/time lags were
explained either in terms of down-scattering of hard photons emitted
in the accretion columns by the colder surrounding plasma
\citep{fal07,cui98,tit02} or by the broader emission pattern of hard
photons Comptonized in the accretion columns with respect to soft
photons emitted from the NS hot spots, which make harder photons to be
seen before the softer ones as the NS rotates
\citep{pou03,gierlinski2005,ibr09}. Above an energy of $6$~keV, the
dependence of the lags observed from IGR J00291+5934 reverses, and the
harder photons ($\sim 100$~keV) arrive later than the softer
($\sim$10-20~keV) ones. In the Comptonization scenario this is
explained by the fact that higher energy photons were up-scattered
more times and took longer to reach the observer \citep[see,
  e.g.,][and references therein]{mfalanga12,fal07,mfalanga11}.

\subsection{Thermonuclear type-I X-ray bursts}
\label{sec:amspburst}
\begin{figure}[t!]
\centering 
\includegraphics[width=\columnwidth]{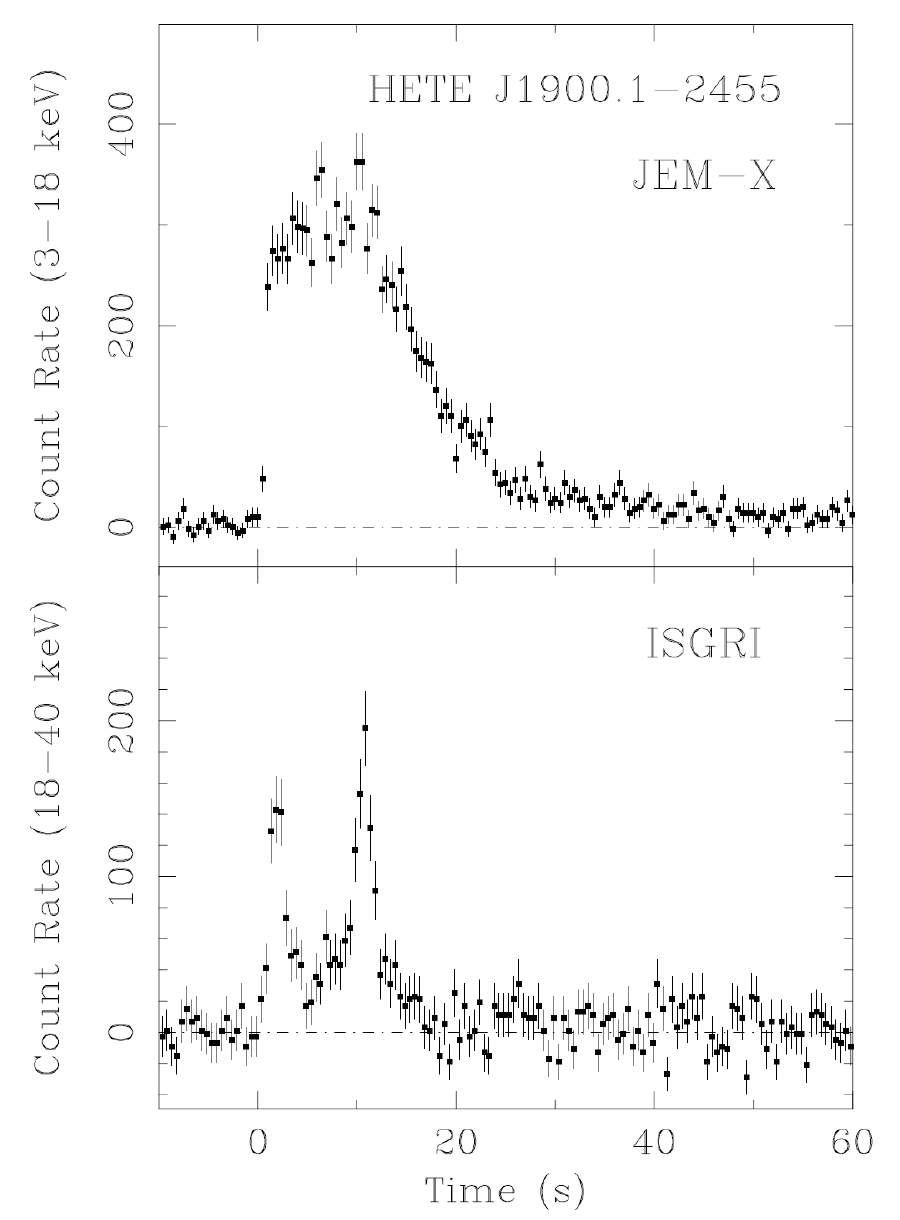}
\caption{An intense {\bf type-I X-ray} burst observed by JEM-X and IBIS/ISGRI from the AMSP HETE~J1900.1-2455 during its outburst in 2005. The upper panel shows the JEM-X light curve (3-18 keV), while the lower panel shows the IBIS/ISGRI light curve (18-40 keV). The double peaked IBIS/ISGRI light curves clearly show the presence of a photospheric radius expansion. Credit: Falanga
et al., A\&A, 464, 1069-1074 (2007), reproduced with permission (ESO).}
\label{fig:inte_burst}
\end{figure}

\begin{figure}[t!]
\centering 
\includegraphics[width=\columnwidth]{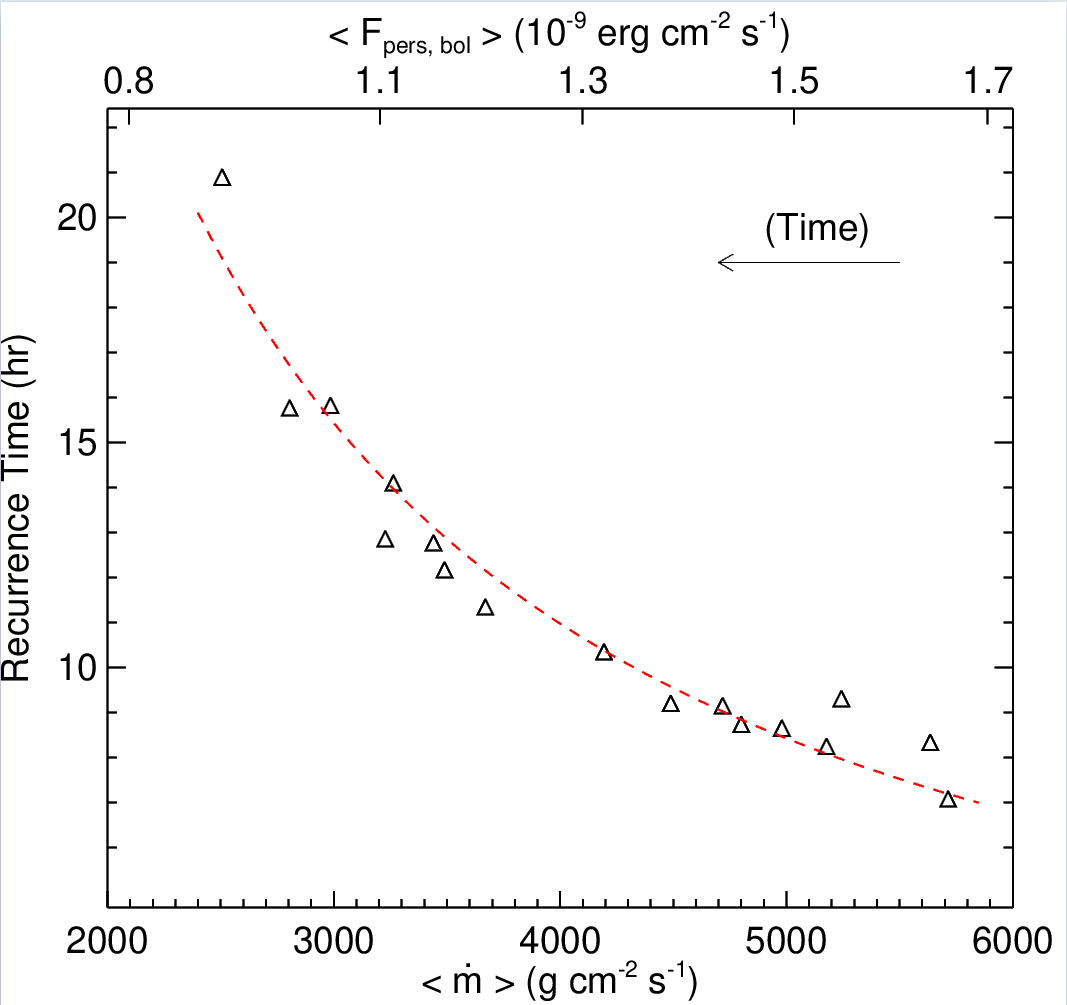}
\caption{An example of the study of the burst recurrence time in the AMSP IGR~J17511-3057 during its 2010 outburst. Triangles in the figure represent the observed burst recurrence times shown as a function of the mass accretion rate per unit area. By using a fit to the data with a power-law model, the recurrence time is found to increase with time roughly as $\langle F_{\rm pers,bol}\rangle^{-1.1}$. Credit: Falanga et al.,
A\&A, 529, A68 (2011), reproduced with permission (ESO).}
\label{fig:inte_burst_rec}
\end{figure}

The {\inte} observing strategy  generally implies a few-day long observations toward a pre-defined region of the sky.  The  {\inte} monitoring programs aimed at the Galactic bulge, as well as dedicated observational campaigns  devoted to specific sources, often offered nearly uninterrupted light curves  of all sources in the field and covering a significantly larger fraction of the outbursts of AMSPs with respect to other telescopes. These light curves were particularly useful to search for thermonuclear type-I X-ray bursts and accurately constrain their recurrence time as a function of the mass accretion rate ($\dot{m}$) inferred from the observed non-burst X-ray flux \citep{lew93}.  Fig.~\ref{fig:inte_burst} shows an example of a particularly intense burst observed by {\inte} from the AMSP HETE~J1900.1-2455 \citep{mf07}, during which the double peaked profile in the IBIS/ISGRI data proved that a photospheric radius expansion took place \citep[see, e.g.,][for a recent review]{galloway17}. On the other hand, the light curves of most of the thermonuclear bursts shown by the AMSPs  observed by {\inte}  were characterized by a fast rise and an exponential decay lasting a few tens of seconds. Such short burst profiles indicate that the ignition most likely occured in presence of  hydrogen-poor material, suggesting that either the accreted material is hydrogen-deficient or that the CNO abundances in AMSPs was slightly higher than the solar value \citep[see, e.g.,][]{falco17b,mfalanga11,mfalanga12,def17,mf07,li18,ferrigno11}. In the case of IGR\,J17511-3057, this suggestion could be strengthened by the fact that the variation of the burst recurrence time as a function of $\dot{m}$ (see Fig.~\ref{fig:inte_burst_rec}) was found to be much  shorter than that predicted in case of helium-ignition models \citep{mfalanga11}. 

\begin{figure*}[t!]
\centering 
\includegraphics[width=13cm]{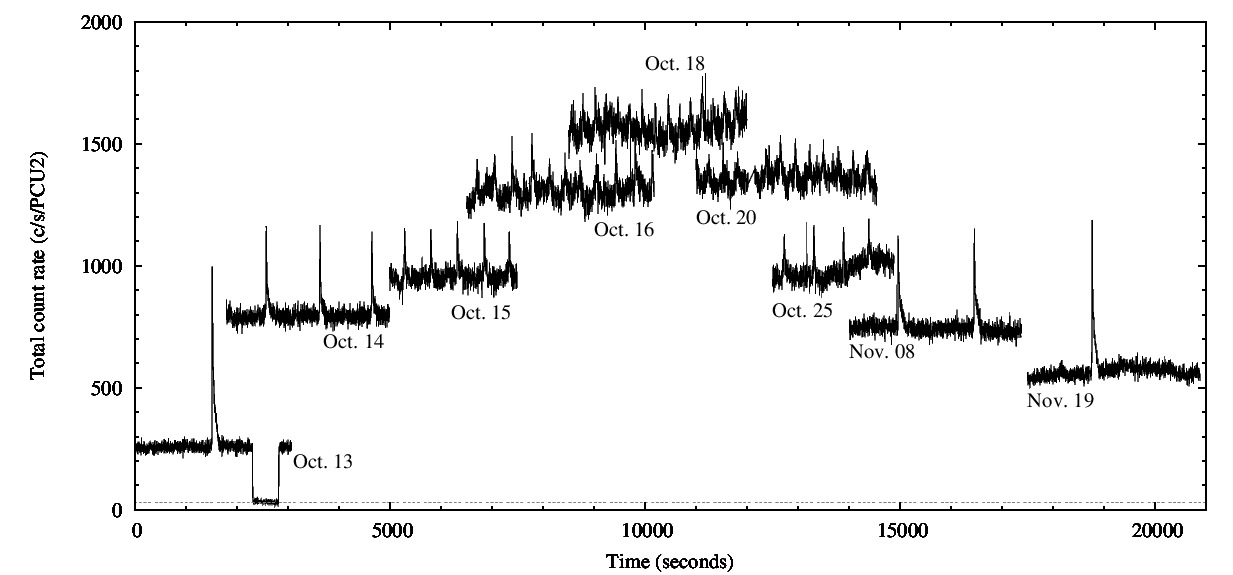}
\caption{The evolution of the thermonuclear burst frequency in IGR\,J17480-2446 during the course of its 2010 outburst. Data were obtained from \rxte\ and the x-axis shows the  time at different phases of the outburst, which evolution of the persistent emission is represented in the y-axis in units of count-rate recorded by the PCU2 on-board \rxte\  Times have been shifted arbitrarily for display purposes. Credit: Linares et al., ApJ, 748, 2 (2012). reproduced
with permission of the AAS.}
\label{fig:J17480}
\end{figure*}

Among the many thermonuclear bursts observed by \inte\ from the more than a hundred of  bursting sources known  \citep[see, e.g.,][for a recent catalogue of the bursts observed by JEM-X]{bursts}, it is worth mentioning here the peculiar case of IGR\,J17480-2446. This source is located in the Globular Cluster Terzan~5 and underwent so far a single bright detectable outburst in 2010 \citep{bor10} which displayed a number of remarkable unique features. First of all, pulsations were clearly detected at a spin period of about 91~ms, making this a unique source linking AMSPs and slower spinning pulsars in LMXBs \citep{str10, pap11, pap12, testa12}. Evolutionary calculations suggested that this binary system might have formed through a close encounter between the NS and its companion within the globular cluster, and that the compact object is only a mildly recycled pulsar as accretion did not begin earlier than a few 10$^7$~yr ago \citep{pat12}. 
Furthermore, the source showed several hundreds of thermonuclear bursts along the outburst. The recurrence time markedly decreased during the rising part of the outburst  (\citep{mot11,linares12}; see Fig.~\ref{fig:J17480}). Close to the outburst peak the frequency of the bursts became so high and their peak flux so close to the level of the persistent emission  that they could no longer be identified by visual inspection in
 the source light curve (\citep{mot11, altamirano12}, see Fig.~\ref{fig:J17480}), but rather emerged as mHz quasi periodic oscillations in the Fourier power density spectrum \citep{rev01}. A careful analysis of the spectral softening of the emission during the burst decay proved unequivocally that they had all a thermonuclear origin \citep{mot11,linares11, chak11}. Although a decrease of the recurrence time with increasing mass accretion rate had been theoretically anticipated \citep{heger}, this was the first time that it was observed in an LMXB in such great detail. This made IGR\,J17480-2446 a unique laboratory to test thermonuclear burning models. All the thermonuclear bursts also displayed burst oscillations at a frequency within a few per cent of the spin rotation of the NS, proving for the first time that millisecond rotational velocities are not required to produce these kind of timing features and ruling out models based on this assumption \citep{cav11}. IGR\,J17480-2446 also displayed a clear evidence of a fast-moving disk wind  with ejection velocity up to $\sim$3000~km~s$^{-1}$, a rare feature in NSs hosted in  LMXB and much more common in systems harboring accreting BHs \citep{miller}. In addition, the source endured an extremely high level of crustal heating during the outburst, which did not seem to have cooled completely even 5.5 years since the end of the outburst \citep{degenaar11, degenaar13, ootes19}. 

Peculiar burst properties have also been observed from SAX J1748.9$-$2021. The burst recurrence time drastically decreased from $\approx 2$ to $\approx 1$ hr as the source underwent an abrupt hard-to-soft state spectral transition (\citep{li18}, see Fig.~\ref{fig:inte_lc2}). The relation between the burst recurrence time and the mass accretion rate  indicated that in both states the bursts were consistent with being produced by a mixture of H and He. 

{\inte} observations of thermonuclear type-I X-ray bursts from LMXBs also played an important role in the study of the properties of the X-ray coronae from the effect of the burst emission on the surrounding accretion flow (see \citep{deg18} and references therein). Stacking of 123 bursts observed by {\inte} from the persistent LMXB  4U 1728-34 revealed a deficit of hard 40$-$80~keV photons compared to the persistent emission due to the enhanced corona cooling caused by the soft burst photons \citep{kaj17}; such a deficit had not been detected before by {\rxte} \citep{ji14}, possibly due to the different response and background contamination. The reader is referred to the article by Sazonov et al. in this volume for more details on bursts of LMXBs.

\section{Transitional millisecond pulsars}
\label{sec:trans}

According to the classical recycling picture \citep{bisnovatyikogan1974,alpar1982,radhakrishnan1982}, the switch on of a millisecond radio pulsar powered by the rotation of its magnetic field should occur only after the Gyr-long mass accretion phase in a LMXB has ceased and the  pulsar wind pressure becomes dominant. However, the possibility that a source could swing between a radio pulsar behaviour and a LMXB regime also over much shorter timescales (days to weeks) had been proposed already a few years before AMSPs were actually discovered  \citep{ste94, campana1998, burderi2001}.  When the X-ray luminosity of a transient LMXB drops below $10^{32}$ erg s$^{-1}$ at the end of an outburst, the magnetospheric radius of a $\sim 10^8$~G NS spinning at a period of a few milliseconds expands beyond the light cylinder radius of the pulsar, $r_{LC}=71.6 (P_s/\mbox{1.5~ms})$~km, where $P_s$ is the pulsar spin period. A radio pulsar powered by the rotation of its magnetic field could then eject the residual disk matter and power the emission of the binary in quiescence.  The discovery that AMSPs were weakly magnetized X-ray transients which dropped to an X-ray luminosity of $10^{31}-10^{32}$ erg s$^{-1}$ during quiescence (see Sect.~\ref{sec:amsplc})  was compatible with such a picture \citep{campana2002}. Furthermore, the optical luminosity of quiescent AMSPs  turned out to be too large to be compatible with reprocessing of such a faint X-ray emission, suggesting it was instead due to reprocessing of the more powerful spin-down energy \citep{bur03}.  The rate at which AMSPs spin down during quiescence \citep{hartman2008} and the fast and complex orbital evolution \citep{hartman2008,disalvo2008} were also very similar to those observed from the so-called redback millisecond radio pulsars \citep{str19,rob13}. However, deep searches for radio pulsations from AMSPs in quiescence were not successful, even when conducted at high radio frequencies (5 and 8 GHz) at which free-free absorption is less important \citep{patruno2017b,burgay2003,iac10}.

\begin{figure*}[t!]
\centering 
\includegraphics[width=13cm]{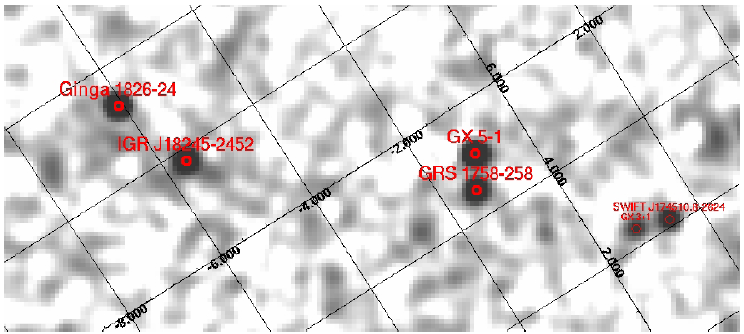}
\caption{Mosaic of the IBIS/ISGRI field around IGR~J18245$-$2452 at the time of its discovery in 2013 \citep{papitto2013}.}
\label{fig:j18245}
\end{figure*} 

The quest for sources switching back and forth between a radio pulsar
and an accretion powered phase finally succeded when the radio pulsar
PSR~J1023+0038 showed indications of a past accretion disk activity
\citep{archibald2009}, and eventually with the discovery of
IGR~J18245-2452, an AMSP that is detected as a radio pulsar during
quiescence \citep{papitto2013}. Dubbed {\it transitional} millisecond
pulsars (see \citealt{campana2018,pap20} for recent reviews), these sources
are crucial to investigate how the interaction between the NS
magnetosphere and the disk in-flow determines the pulsar emission
regime. {\inte} played a crucial role in the recent discovery of this
class mainly thanks to its all-sky monitoring capabilities. On one
hand it detected the onset of the outburst of IGR~J18245-2452
\citep{eckert2013}; on the other, its catalogue included steady and
relatively faint hard sources which were later identified as candidate
transitional millisecond pulsars through multi-wavelength follow-up
observations \citep{strader2016,cotizelati2019}.

\subsection{Swinging between accretion and rotation-powered states, IGR~J18245-2452}
\label{sec:tmspj18245}
In 2013, IBIS/ISGRI discovered the transient IGR~J18245$-$2452 in the globular cluster M28 \citep[][see Fig.~\ref{fig:j18245}]{eckert2013}.  The luminosity observed from the transient ($3\times10^{36}$~erg s$^{-1}$ at a distance of 5.5~kpc) suggested that it was powered by mass accretion in a binary system. Subsequent observations of type-I X-ray bursts both by {\inte} \citep{def17} and {\swift} \citep{papitto2013} identified the compact object in the binary as a NS. Finally, a coherent X-ray periodicity at the 3.9~ms spin period of the NS was detected thanks to high-time resolution {\xmm} observations; the spin and orbital parameters of this newly-discovered AMSP were the same as those of a radio pulsar detected a few years before when the X-ray source was in quiescence, so unveiling the transitional nature of the system \citep{papitto2013}.

\begin{figure}[t!]
\centering 
\includegraphics[width=\columnwidth]{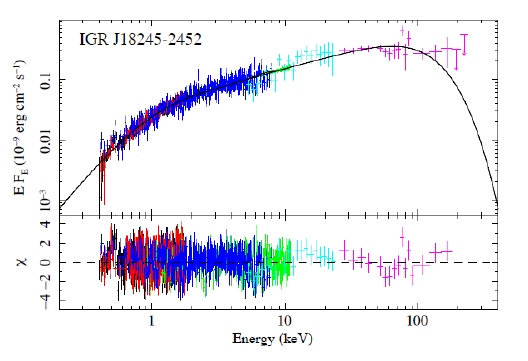}
\caption{The unfolded absorbed broad-band spectrum of IGR~J18245$-$2452 observed during the 2013 X-ray outburst, fitted with a Comptonization  {\sc compps} model (black line). Data points from {\xmm}/RGS (red and black points, 0.4$-$1.8~keV), {\xmm}/EPIC-pn (green points, 0.9$-$11~keV), {\swift}/XRT (blue points, 0.4$-$8~keV), {\inte}/JEM-X (light blue points, 5$-$25~keV) and  {\inte} IBIS/ISGRI (magenta points, 22$-$250~keV) were included. The bottom panel shows residuals with respect to the best fit model. Credit: De Falco
et al., A\&A, 603, A17 (2017), reproduced with permission (ESO).}
\label{fig:j18245spec}
\end{figure}

\begin{figure*}[h!]
\centering 
\includegraphics[width=11cm]{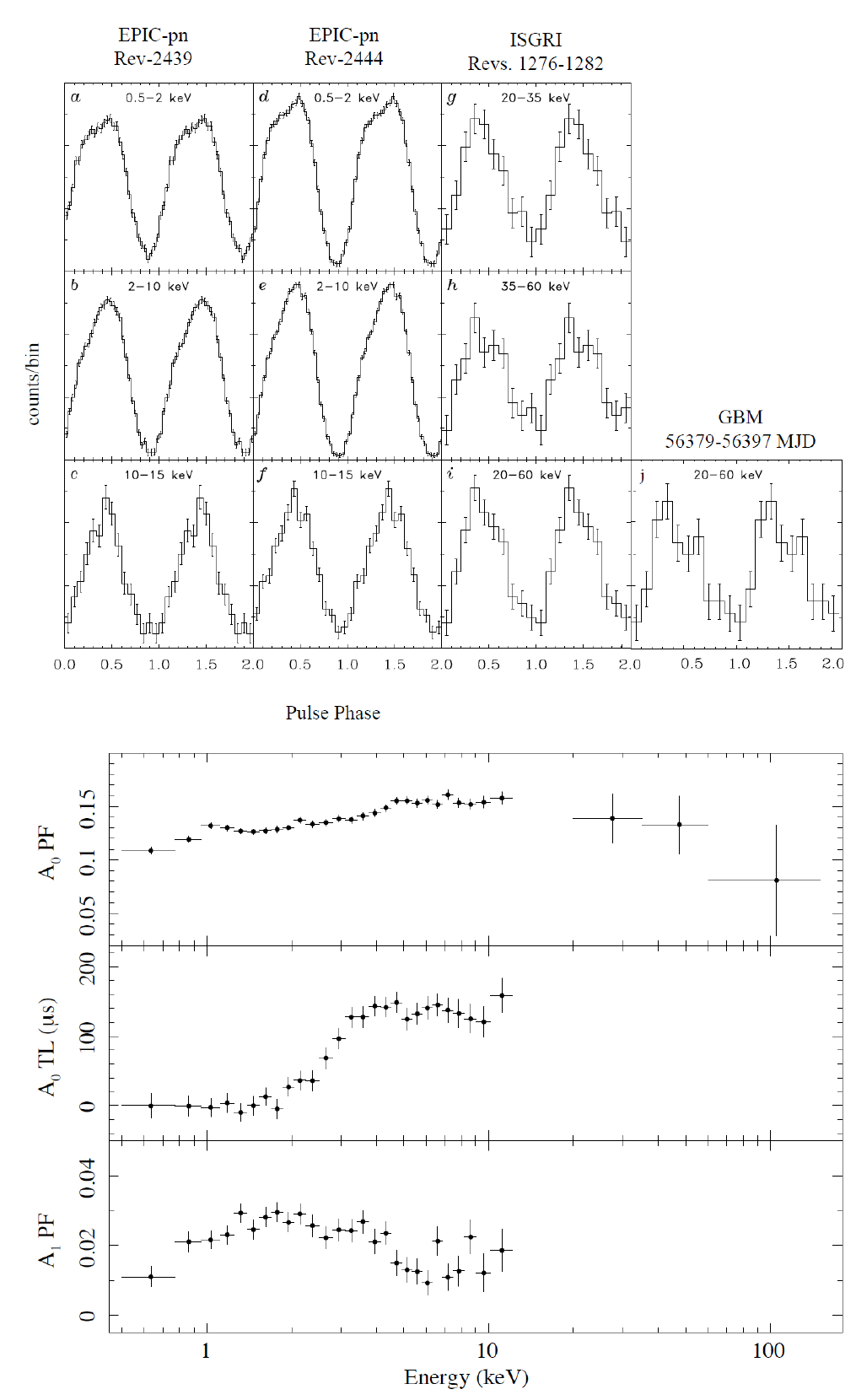}
\caption{Top panel: pulse profiles of IGR~J18245$-$2452 observed by {\xmm}/EPIC-pn (panels a)-f), {\inte} IBIS/ISGRI (panels g)-i) and {\it Fermi}/GBM (panel j);  bottom panel, from top to bottom: pulsed fraction of the first harmonic, hard phase/time lags, and pulsed fraction of the second harmonic. Credit: De Falco et al.,
A\&A, 603, A17 (2017), reproduced with permission (ESO).}
\label{fig:j18245pulse}
\end{figure*} 

IBIS/ISGRI and JEM-X observations  measured the properties of this transitional pulsar in the accretion phase  and allowed to put it in the context of AMSPs \citep{def17}. The observed spectral energy distribution (see Fig.~\ref{fig:j18245spec}) was dominated by a power law component with a photon index of $\Gamma=1.32(1)$ and a cut-off energy of $122^{+21}_{-16}$~keV. Interpreted in terms of Comptonization of soft photons radiated from the NS hot spots, the spectral energy distribution measured from IGR J18245$-$2452 was the hardest among AMSPs.
 X-ray pulsations were detected by IBIS/ISGRI up to 60 keV, with a pulsed fraction of $\sim10\%$, compatible with the values detected at low energies (see the top panel of Fig.~\ref{fig:j18245pulse}). The high energy pulses of IGR J18245$-$2452 lagged behind the pulses detected at soft X-rays by up to 150 $\mu$s (see the middle panel of Fig.~\ref{fig:j18245pulse}).  This was normally not observed in other AMSPs (see Sect.~\ref{sec:amsppulse}) and was ascribed to a peculiar energy dependence of the emission pattern of the hot spots on the NS surface \citep{def17}. Type-I X-ray bursts observed by {\inte} and {\swift} were all powered by ignition of Helium.

\subsection{X-ray sub-luminous transitional millisecond pulsars}

\begin{figure*}[h!]
\centering 
\includegraphics[width=15cm]{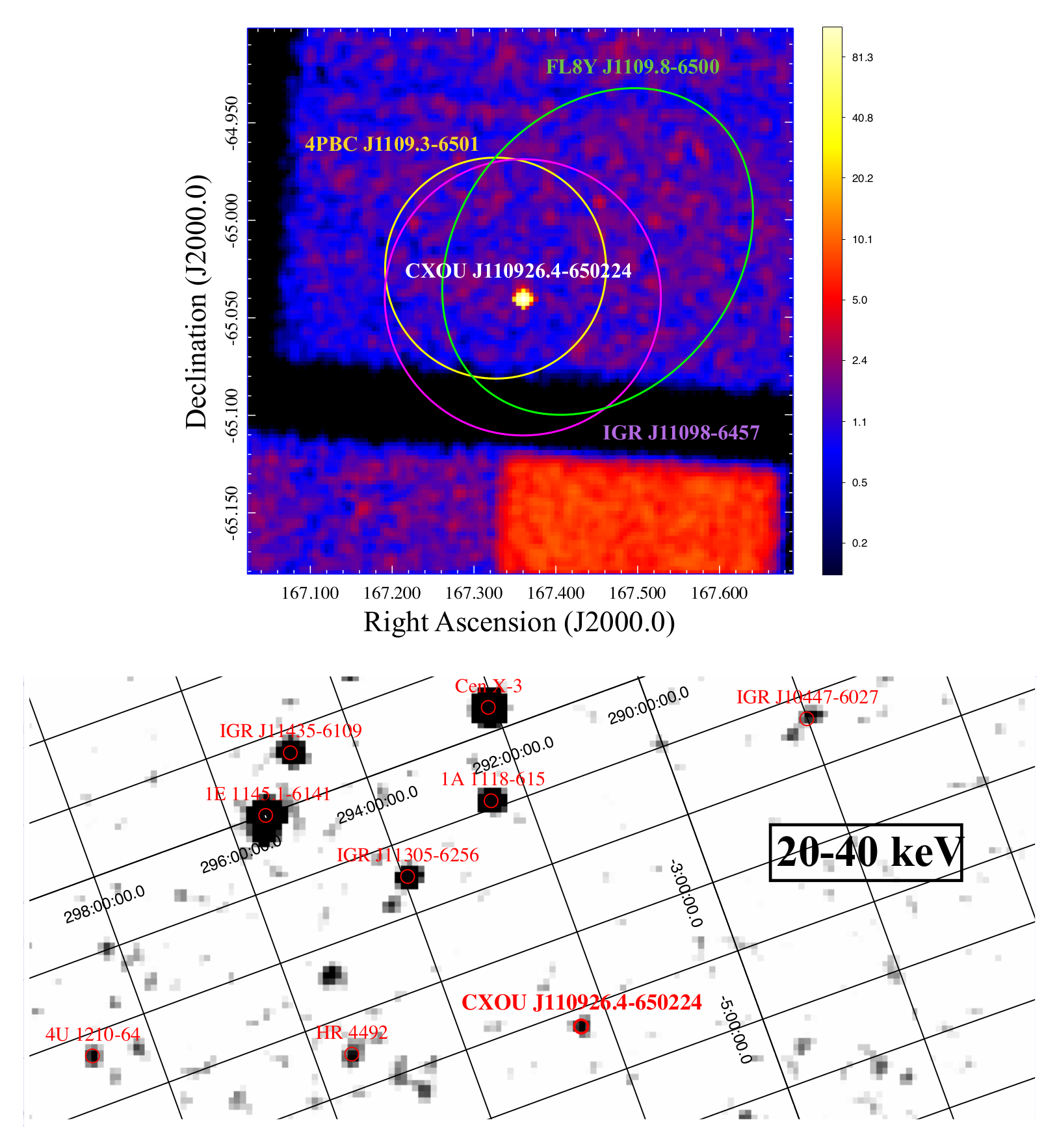}
\caption{Top panel: {\it Chandra}/ACIS-I 0.3$-$8~keV image of the field around CXOU~J110926.$-$650224 with the error circles of the counterparts from {\inte} (IGR J11098$-$6457, 20$-$40~keV, magenta line) {\swift}/BAT (4PBC~J1109.3$-$6501, 15$-$150~keV, yellow line) and {\it Fermi}/LAT (FL8Y~J1109.8-6500, 100~MeV$-$300~GeV, green line) catalogues.  Bottom panel: {\inte} IBIS/ISGRI 20$-$40keV mosaic image of the field around IGR J11098$-$6457. Credit: Coti Zelati et al., A\&A, 622, A211
(2019), reproduced with permission (ESO).}
\label{fig:1109}
\end{figure*} 

The month-long outburst observed in 2013 from IGR J18245$-$2452 has been the only bright accretion event observed from a transitional millisecond pulsar so far. Other transitional millisecond pulsars, such as PSR J1023+0038 \citep{archibald2009} and XSS J12270-4859 \citep{bassa2014}, have persisted for years in an accretion disk state characterized by a much fainter X-ray luminosity ($L_X\simeq5\times10^{33}$ erg s$^{-1}$), variable among two roughly constant levels (dubbed {\it high} and {\it low} modes) and frequent flares \citep{demartino2010,lin14,pat14,bogdanov2015}.  This peculiar accretion disk state was also characterized by a variable, bright continuous radio emission \citep{del15,bogdanov2018} and by an unexpected $\gamma$-ray ($\sim$~GeV) emission with a power roughly comparable to that observed in the X-ray band \citep{sta14,torres2017}. X-ray pulsations were detected only in the {\it high} mode \citep{archibald2015,papitto2015}, simultaneously to unexpectedly bright optical pulses \citep{ambrosino2017,papitto2019}. 

The ability of IBIS/ISGRI in detecting faint, quasi persistent hard X-ray sources was crucial to identify such transitional pulsars in this enigmatic state. The case of XSS J12270-4859 is illustrative. A program of optical spectroscopy of  unidentified {\inte} sources showed the presence in its spectrum of broad, double-peaked emission lines originating from an accretion disk \citep{masetti2006}. While such a spectrum hinted at a cataclysmic variable, the presence of a $\gamma$-ray counterpart detected by the {\it Fermi}/LAT instrument suggested an atypical low-mass X-ray binary instead \citep{demartino2010,hil11,demartino2013}. The IBIS/ISGRI spectrum was described by a power law with a hard photon index ($\Gamma=1.3$) extending up to 100 keV without a detectable cut-off \citep{demartino2010}. The continuous detection by IBIS/ISGRI showed that the source could persist in such a state for at least a decade, possibly powered by a millisecond pulsar that ejects disk mass through the propeller effect \citep{papitto2014a,pap15b} or accreting material at a very low rate \citep{dang12,boz18}. The transitional nature of the source was eventually demonstrated when its disk disappeared \citep{bassa2014} and it  switched on as a radio pulsar \citep{roy2015}. 

Cross referencing the {\inte} IBIS/ISGRI catalogue with that of the Large Area Telescope (LAT; 20 MeV - 300 GeV) aboard the {\it Fermi Gamma-ray Space Telescope} \citep{fermi2019} turned out to be one of the most effective techniques to identify strong candidate transitional millisecond pulsars in such a faint and peculiar accretion disk state (see Fig.~\ref{fig:1109}), and allowed the identification of the candidates IGR J04288$-$6702 \citep{strader2016} and  IGR J11098$-$6457 \citep{cotizelati2019}.  Follow-up observations of these two sources at optical and soft X-ray energies hinted at similar time variability properties as those observed from the other known transitional millisecond pulsars. However, their transitional nature is yet to be firmly established, waiting for a switch to the radio pulsar state hopefully in the near future.

\section{Rotation-powered pulsars}
\label{sec:radiopulsars}
\newcommand{\gr}{$\gamma$}

Before the launch of {\inte}, highly magnetized
($B\simeq10^{11}-10^{12}$~G) rotation-powered pulsars were already
known to emit steadily from radio frequencies up to high-energy
$\gamma$-rays. The Compton Gamma-Ray Observatory ({\it CGRO}) had
increased the number of detected high-energy ($E \ge$100 MeV)
$\gamma$-ray pulsars to a total of eight (see \citealt{thompson1997}
for a review). However, the timing signatures and spectral shapes in
the energy band from $\sim$ 20 keV to $\sim$ 5 MeV (or higher) were
only measured for three pulsars: the Crab pulsar (PSR B0531+21), PSR
B1509-58 and the Vela pulsar (PSR B0833-45).  This very small sample
exhibited very different characteristics. The Crab and Vela pulsars
showed double-peaked pulse profiles, while PSR B1509-58 showed a broad
single pulse. The latter had a spectral shape peaking in luminosity at
MeV energies, vastly different from that of the Crab
\citep{kuiper1999}. The Vela pulsar was only marginally detected at
hard X-ray/soft $\gamma$-ray energies but appeared to be the strongest
pulsar at GeV energies, where it reached maximum luminosity with a
spectral shape completely different from both the Crab and PSR
B1509-58 \citep{hermsen1994}.

The high-energy non-thermal emission from highly magnetized rotating
NSs was believed to originate from particle acceleration along the
open magnetic field lines in the pulsar magnetosphere. The different
competing models could be grouped in two categories, the so-called
Polar Cap (PC) models and Outer Gap (OG) models. In the PC scenario
charged particles (mainly $e^{+}e^{-}$) are accelerated along the open
field lines in the vicinity of the magnetic poles. Subsequent cascade
processes (starting with curvature radiation or inverse Compton
scattering) give rise to the emergent high-energy spectrum
(e.g. \citealt{daugherty1994}). In the OG scenario the acceleration of
charged particles and production of high-energy radiation takes place
in charge depleted gaps between the null-charge surface and the light
cylinder above the last closed field lines
(e.g. \citealt{cheng1986a,cheng1986b}). It was also predicted that due
to their quick rotation, millisecond pulsars should be fairly strong
high-energy $\gamma$-ray emitters \citep{srinivasan1990}.

With the launch in June 2008 of  {\it Fermi}, the LAT produced ground-breaking results at high-energy $\gamma$-rays. The number of detected $\gamma$-ray pulsars increased from eight to $\sim$ 250 (status in 2019), including 90 millisecond pulsars (MSPs). These numbers can be compared to the seven pulsars securely detected above 100 MeV with the Energetic Gamma Ray Experiment Telescope (EGRET) aboard CGRO and the single recycled MSP J0218+4232, likely detected above 100 MeV \citep{kuiper2000}. The number of $\gamma$-ray sources totals more than 5000 in the fourth {\it Fermi} catalog \citep{fermi2019}, including 239 identified pulsars and many candidate rotation-powered pulsars. These results also stimulated  major developments in theoretical modelling of the high-energy emissions in pulsar magnetospheres. The new emission characteristics, e.g. spectral turnover at GeV energies and pulse morphology, favoured OG models over PC models \citep{harding2005} and Slot Gap models \citep{dyks2003,muslimov2004}. However, recent global particle-in-cell simulations of pulsar magnetospheres reveal that most particle acceleration occurs in and near the current sheet beyond the light cylinder and the separatrices \citep{che14,mochol2015,cerutti2016,philippov2018,brambilla2018,kalapotharakos2018}.

{\inte} promised for the first time fine imaging and accurate localization of hard-X-ray/$\gamma$-ray sources, with good timing and good sensitivity over the broad energy range from 3 keV to $\sim$ 10 MeV. However, the predictions on the expected results for rotation-powered pulsars over this energy window were very uncertain. They relied on uncertain interpolations performed on a handful of radio pulsars between the spectra of pulsed emission measured at higher $\gamma$-ray energies and at soft X-ray energies, or on extrapolations of the spectra measured only at lower X-ray energies. As expected, the strong Crab pulsar emission could be used as a calibration source, and studied in more detail at hard X-rays. However, the next-in-flux known hard X-ray pulsar, PSR B1509-58,
whose hard X-ray emission had been studied earlier using the instruments aboard {\it CGRO} \citep{kuiper1999} and the Italian/Dutch mission {\it BeppoSAX} \citep{cusumano2001}, was   already a factor 30-50 times weaker than the Crab in the 20-100 keV band. The Vela pulsar was even fainter, with a hard X-ray luminosity  $\sim$ a thousand  times lower than the Crab. Using data from the {\rxte}, hard X-ray timing and spectral properties had been also reported for the Crab-like pulsar in the Large Magellenic Cloud (LMC), PSR B0540-69 \citep{deplaa2003}, which turned out to be $\sim$250 times weaker than that of the Crab in the hard X-ray domain. For the recycled rotation-powered pulsars, the MSPs, the prospects for detections with {\inte} were even smaller. Three MSPs, PSR B1821-24, PSR J0218+4232, and  PSR B1937+21, had been reported to emit non-thermal emission at X-ray energies with very hard spectra with power-law indices $\Gamma \sim$ 1.1. X-ray pulsations were detected up to 20 keV for PSR B1821-24 \citep{rots1998} and PSR J0218+4232 \citep{kuiper2004}, and up to 25 keV for PSR B1937+21 \citep{cusumano2003}. However, the reported fluxes were even more than three orders of magnitude weaker than that of the Crab at 20 keV.

Given the expected low {\inte} count rates of the weak pulsed emission, it is no surprise that in the first decade of the {\inte} mission, most analyses did not address the pulsed emission but exploited the imaging capabilities in analyses of the total emission from pulsars and their Pulsar Wind Nebulae (PWNe). Only later, first detections at hard X-rays were reported and the total spectra (pulsar + PWN) discussed.
The detection of  emission up to 200 keV from PSR B1509-58 in SNR MSH 15-52 \citep{sturner2004} was followed by the presentation of the spatial and spectral properties of just the unpulsed emission in the $20-200$ keV band \citep{forot2006}.
PSR B0540-69 / SNR 0540-693 in the LMC  was first detected up to 100 keV  \citep{gotz2006}, later up to 200 keV, while significant pulsations with a high duty cycle were visible up to 100 keV \citep{slowikowska2007}.
For PSR J1617-5055 near RCW 103 ($18-60$ keV \citep{landi2007}) and later PSR J1811-1925 in G11.2-0.3 (up to $\sim$ 200 keV\citep{dea08b}) the total spectra could be discussed for the first time. In addition, with {\inte} IBIS/ISGRI PSR J0537-6910 was detected in the $20-60$ keV band in a very deep survey of the LMC region \citep{grebenev2013}.

The above summary indicates that very long {\inte} exposures were
required to increase the sample of pulsars for which pulsed emission
could be detected at hard X-rays. Similar to the case of AMSPs (see
Sect.~\ref{sec:amsppulse}), the low pulsed-count rates in the {\inte}
window required timing analyses to rely on pulsar ephemerides
determined in radio monitoring observations, or in the X-ray band
below $\sim$10 keV where instruments such as the PCA aboard {\it RXTE}
were sufficiently sensitive to allow for measurements of pulsar
ephemerides. The latter was required when the pulsars were not
detected in the radio band. Once these timing solutions were
determined at lower energies, phase folding of the event arrival times
measured with the {\inte} instruments could be performed to search for
the pulsed signals at hard X-rays. A complete high-energy overview of
the soft $\gamma$-ray pulsar population was published with a catalog
containing 18 rotation-powered (non-recycled) pulsars with pulsed
emission detected in the hard X-ray band above 20 keV
\citep{kuiper2015}. Surprisingly, most of these pulsars were not
detected by {\it Fermi} at high-energy $\gamma$-rays. This catalog and
the characteristics of this sample will be addressed at the end of
this section. First, the {\inte} results from studies of the Crab
pulsed emission will be presented, followed by the important results
from polarization measurements.

\subsection{The Crab pulsar}
Early in the mission, the first {\inte} results were published for the archetypical Crab pulsar with a verification of the absolute timing capabilities of all high-energy instruments  \citep{kui03,brandt2003}, JEM-X, IBIS/ISGRI and SPI. It was shown that the X-ray main pulse was leading the radio pulse in phase by 285 $\pm$ 12$\,\mu$s (IBIS/ISGRI) and 265 $\pm$ 23$\,\mu$s (SPI) \citep{kui03} (statistical errors only), values that are more accurate than those reported earlier. Using six years of SPI telescope data (total exposure $\sim$4 Ms) comparable values (275 $\pm$ 15\,$\mu$s) were found for the hard X-ray band ($20-100$ keV; \citealt{molkov2010}). More interestingly, it was shown that the delay between the radio and X-ray signals varies in the $20-300$ keV range. Namely, the delay was reported to be 310 $\pm$ 6$\,\mu$s in the $3-20$ keV soft X-ray band from an analysis of {\it RXTE} data \citep{molkov2010}.

A coherent high-energy picture of the Crab nebula and pulsar spectra
from soft X-rays up to high-energy $\gamma$-rays had been published
shortly before the {\inte} launch \citep{kuiper2001}, including a
pulse-phase-resolved spectral analysis performed in seven phase slices
over the 0.1 keV$-$10 GeV energy band. In this high-energy picture of
the Crab, data from the four narrow-field instruments aboard {\it
  BeppoSAX} (LECS, MECS, HPGSPC and PDS) covered energies up to 300
keV (see \citealt{kuiper2001} for references). At higher $\gamma$-ray
energies, data were used from COMPTEL and EGRET aboard {\it
  CGRO}. Early in the {\inte} mission, an accurate phase-resolved (now
in 50 bins) spectral analysis for the Crab pulsar over the energy
range 3$-$500 keV was achieved with multiple Crab calibration
observations with JEM-X, IBIS/ISGRI \& PICsIT and SPI, characterizing
in detail the curved spectral shape over this energy range
\citep{mineo2006}. The combination of {\inte} timing and spectral
results with those from the previous {\it BeppoSAX} and {\it CGRO}
missions were input for a multi-component model for the broad-band
emission of the Crab pulsar from the optical band to high-energy
$\gamma$-rays \citep{massaro2006}.

\subsection{Joint optical - $\gamma$-ray polarisation measurements with {\inte}}

Although a relatively weak radio pulsar and  a strong emitter at X and $\gamma$-ray energies, the Crab pulsar is  the brightest of all the known pulsars at optical wavelengths and consequently has been extensively studied. The pulsar's spin-down energy powers its surrounding nebula, which radiates at all electromagnetic frequencies from the radio band to TeV $\gamma$-rays \citep{hester2008}. Our understanding of the high-energy emission process in pulsars is still very incomplete. However, polarisation observations can begin to unravel this conundrum through geometric considerations. Harding and  Kalapotharakos \citep{harding2017} have modelled the high-energy polarised emission from optical to $\gamma$-ray wavelengths. They predicted a polarisation degree and polarisation angle which depends upon location and specifically whether the radiation originates from inside or outside the pulsar's light cylinder. Matching the polarisation profile can give a unique restriction on the location of the production of high-energy emission \citep{mcdonald2011}.

The Crab nebula was one of the first objects outside of the Solar System to have detectable X-ray polarisation \citep{weisskopf1978}. This was followed by the first detection of $\gamma$-ray polarisation using {\inte} \citep{dean2008, forot2008}.  Optical polarised emission from the nebula and pulsar have
been described by a number of authors \citep{smi88, 
 slowikowska2009, moran2012, moran2013, moran2016}. Phase-resolved observations  showed a change in
polarisation consistent in shape with a beam of synchrotron radiation coming from both poles of an orthogonal rotator.  S\l{}owikowska et al.  \citep{slowikowska2009,slo12} found the presence of a highly polarised continuous component which most likely corresponds to the nearby bright synchrotron emitting knot \citep{moran2013}, also known as inner-knot, located 0.65 arcsec to the southest of the pulsar \citep{hes95}. 

Optical and X-ray observations showed spatial and some flux variability of the inner nebula \citep{bietenholz2001}.
Indeed this had been observed in some early optical studies \citep{scargle1969}. However, the  flux from the whole nebula was expected
to be constant at the level of a few percent \citep{kirsch2005, weisskopf2010} and as such was often used as a standard candle calibration source. However since 2008, strong $\gamma$-ray flares have been observed at a rate of about 1 per year  by the {\it Agile} and {\it Fermi} $\gamma$-ray telescopes \citep{abdo2011, tavani2011, striani2013}. Around these $\gamma$-ray flaring events, there were no associated changes seen in the near-IR or X-ray fluxes \citep{weisskopf2013}.

\begin{figure}[t!]
\centering
\includegraphics[width=\columnwidth]{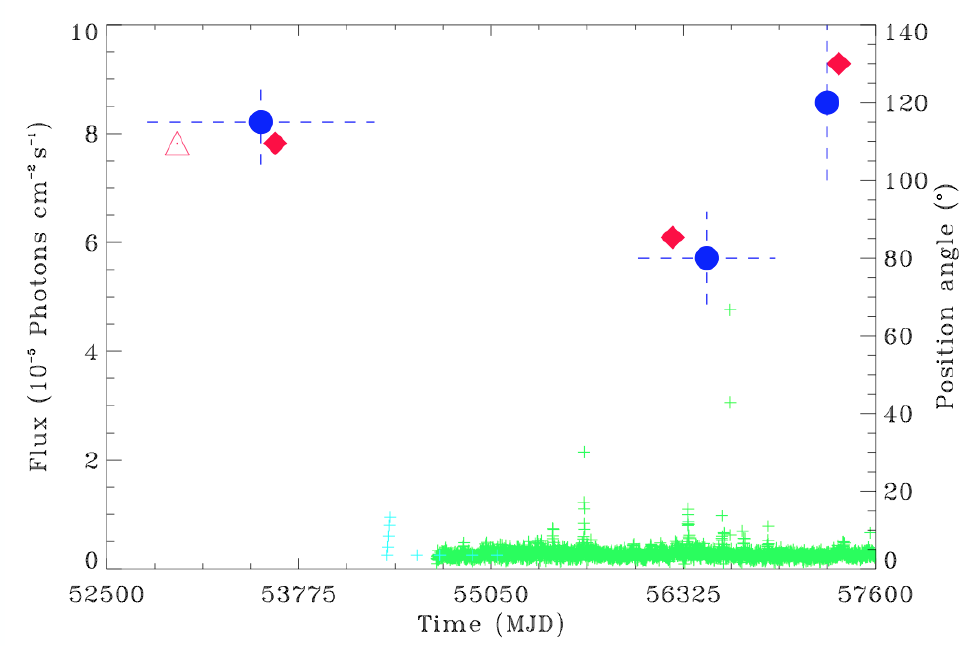}
\caption{Galway Astronomical Stokes Polarimeter (GASP,
  \citealt{col13}, open triangle and filled diamonds) and {\inte}
  observations of the polarisation position angle of the Crab (blue
  filled circles; \citealt{moran2016, oconnor2018}). Green and cyan
  points show the flux observed by {\it Fermi} and {\it AGILE},
  respectively.  More recent Astrosat measurements
  \citep{vadawale2018} in the 0.1-0.38 MeV band indicate a
  polarisation angle of $(143.5 \pm 2.8)^{\circ}$ consistent with the
  trend observed by GASP and {\inte}. Credit: Moran et al., MNRAS,
  456, 2974 (2016), reproduced with permission of Oxford University
  Press on behalf of the Royal Astronomical Society. } \label{pol}
\end{figure}

Since the original {\inte} $\gamma$-ray observations in 2008, the Crab has been observed twice a year improving the statistics for determining the  $\gamma$-ray polarisation \citep{jou19}. This has allowed for comparisons between optical  and MeV $\gamma$-ray polarisation. In particular Moran et al. \citep{moran2016} looked for any correlated changes in the polarisation at the two wavebands. Changes in the  polarisation angle seemed to be correlated, albeit only at the 2.5 $\sigma$ level (see Fig.~\ref{pol}). Whether this is indicative of reconnection or other events associated with $\gamma$-ray flares is not proven. Other suggestions include magneto-Bremsstrahlung emission which explain the lack of variability at lower photon energies \citep{weisskopf2013}.  More correlated observations simultaneous with a flare are required to address this problem, either looking at flux and polarisation changes in the nebula and/or in the immediate vicinity of the pulsar.

\subsection{The soft $\gamma$-ray pulsar catalog}

\begin{figure}[t!]
     \includegraphics[width=\columnwidth]{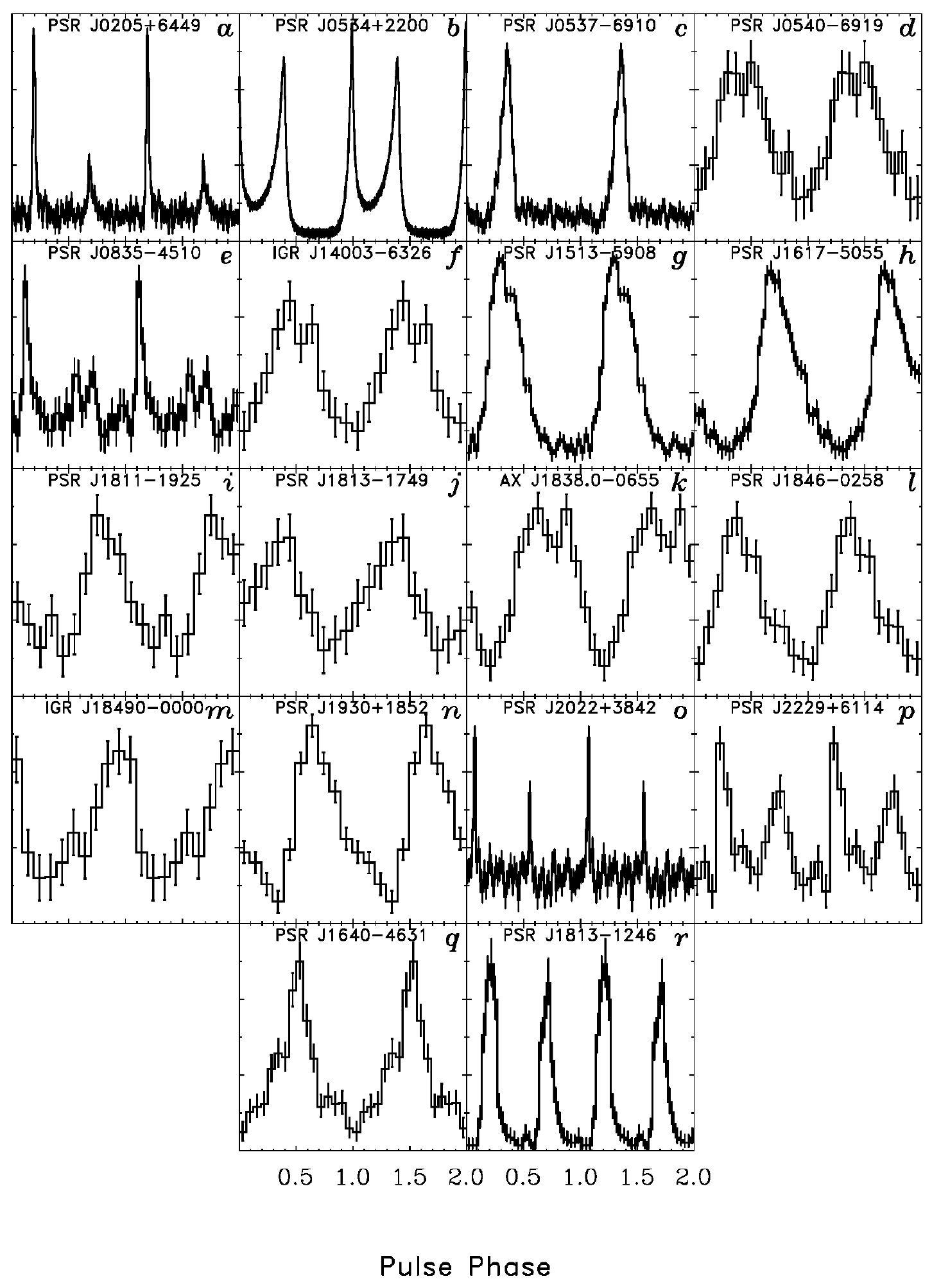}
     \caption{\label{he_psr_morph}The pulse shapes of all soft \gr-ray pulsars in the catalog \citep{kuiper2015}. The profiles are measured either with {\inte} IBIS/ISGRI or RXTE PCA or HEXTE. Exceptions are PSR J1640-4631 ({\it NuSTAR}) and PSR J1813-1246 ({\xmm} EPIC). Credit: Kuiper \& Hermsen, MNRAS, 449, 3827 (2015), reproduced with
permission of Oxford University Press on behalf of the Royal Astronomical
Society.}
\end{figure}

\begin{figure*}[htbp]
  \includegraphics[width=2.1\columnwidth]{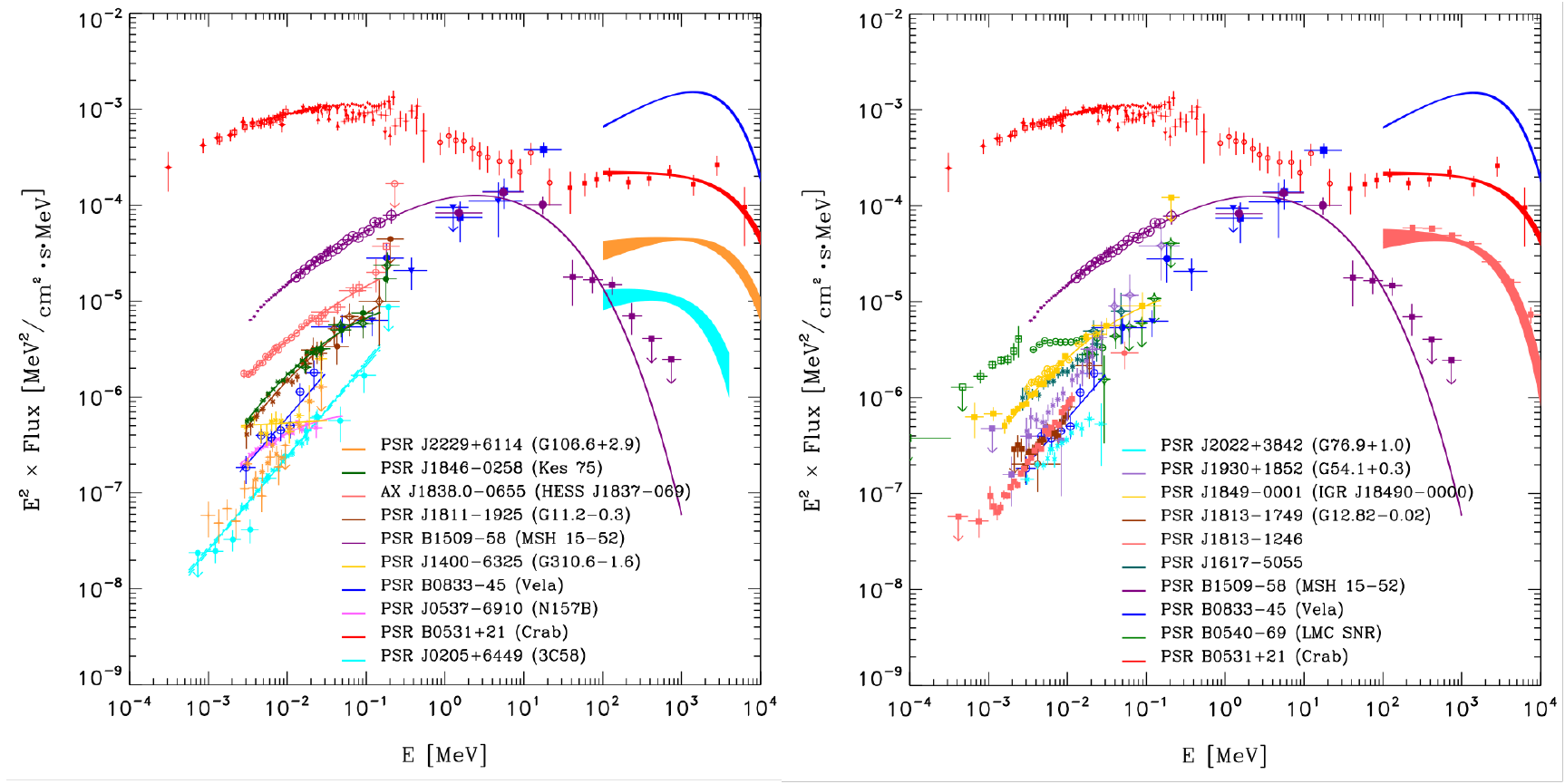}
     \caption{The high-energy {\it pulsed} emission spectra for 17 of the 18 detected soft $\gamma$-ray pulsars from 0.1 keV $-$ 10 GeV  summarized in 
     two panels. The spectra of PSR B0531+21 (Crab), PSR B0833-45 (Vela) and PSR B1509-58 are shown in both panels for reference purposes. The pulsed spectrum of PSR J1640-4631 is not shown, because its {\it total} spectrum has been reported up to hard X-rays but its {\it pulsed} spectrum just up to 25 keV. Credit: Kuiper \& Hermsen, MNRAS, 449,
3827 (2015), reproduced with permission of Oxford University Press on behalf
of the Royal Astronomical Society).}
     \label{spectralcompilation}
\end{figure*}

Despite the weakness of the pulsar signals in the {\inte} hard X-ray/soft $\gamma$-ray band, significant progress was achieved when the X-ray observatories {\it Chandra}, {\rxte}, {\xmm}, {\it Suzaku}, {\inte} and later {\it NuSTAR}, discovered weak energetic point sources at soft/medium X-rays ($0.1-10$ keV) in young supernova remnants, often detected at radio frequencies or in location-error boxes of unidentified {\it CGRO} EGRET/{\it Fermi} LAT ($\geq100$ MeV), {\inte} IBIS/ISGRI (20-300 keV) or {\it H.E.S.S.}/{\it VERITAS}/{\it MAGIC} ($\geq$30/100 GeV) sources. The pulsed signals of these new point sources could be identified at soft X-rays or in the radio band. Once the timing solutions were determined below 10 keV, phase folding of the event arrival times measured with {\rxte} PCA and HEXTE and {\inte} IBIS/ISGRI could be performed to search for the pulsed signals above 20 keV. In addition, the imaging capabilities of IBIS/ISGRI allowed  the detection and spectral characterization of the total emission (pulsar plus PWN), often up to energies above 100 keV.
However, very long exposures were required to obtain secure detections, collecting over many years of exposure, each time a source was in the field-of-view of {\inte} and/or {\rxte}. This approach led to an {\inte}/{\it RXTE} soft $\gamma$-ray pulsar catalog containing 18 pulsars for which non-thermal pulsed emission has been securely detected at hard X-rays/soft $\gamma$-rays above 20 keV \citep{kuiper2015}. That paper summarizes
the history of the detection of each pulsar in different bands of the electromagnetic spectrum, and the X-ray timing and spectral characteristics. Below, we will present  the role of {\inte} in the discovery and/or characterization of a few pulsars, followed by a comparison of the sample of hard X-ray/soft $\gamma$-ray pulsars with those in the second {\it Fermi} pulsar catalog \citep{abd13}.

\subsubsection{IGR J18490$-$0000, IGR J14003$-$6326 and IGR J11014$-$6103 from {\it INTEGRAL} sources to pulsars}

Three pulsars were originally discovered as X-ray point sources by {\inte} in deep IBIS/ISGRI mosaic images, namely IGR J18490$-$0000/PSR J1849-0001 \citep{molkov2004}, IGR J14003$-$6326/PSR J1400-6325 \citep{keek2006} and  IGR J11014-6103/PSR J1101-6101 \citep{bird2010}. Follow-up observations outside the {\inte} band led to their identification as rotation-powered pulsars. 

Multiple follow-up observations of IGR J18490$-$0000 revealed the presence of a TeV source with HESS, a soft X-ray counterpart with {\it Swift}-XRT and {\it XMM-Newton} and, finally \citep{gotthelf2011},  the 38.5-ms X-ray pulsation and a weak PWN with {\it Chandra} (for further references and derived (hard) X-ray characteristics see \citep{kuiper2015}), while no radio counterpart was detected with the GMRT. The discovery of IGR J14003$-$6326 triggered similar follow up observations, notably with {\it Chandra}, revealing the presence of a point source plus a PWN and a SNR \citep{renaud2010}. The latter reference also reported the detection of a 31.2-ms pulsar with {\it RXTE} PCA. Improved (hard) X-ray characteristics are given with the catalog \citep{kuiper2015}. Both these new hard X-ray pulsars were not detected by {\it Fermi} LAT. For the strongest of the two (PSR J1849-0001) one can conclude that the non-detection by the LAT means that maximum luminosity is reached at MeV energies.

IGR J11014-6103 and its surroundings were studied in great detail in X-rays, first by using all available archival X-ray data \citep{pavan2011} and subsequently with a dedicated {\it Chandra} observation \citep{pavan2014}. It was shown that the putative pulsar counterpart is moving away from SNR MSH 11-61A, located at 11 arcmin from the X-ray point source. The latter was shown to have a hard spectrum in the $2-10$ keV band (power-law photon index $\Gamma= 1.1 \pm 0.2$). Finally, {\it XMM-Newton} observations revealed the X-ray pulsations (62.6 ms; \citealt{halpern2014}) and allowed the characterisation of the pulsed-emission X-ray spectrum up to 10 keV \citep{kuiper2015}. This pulsar is an excellent candidate to be detected at higher X-ray energies as a soft $\gamma$-ray pulsar.

\subsubsection{AX J1838.0-0655 / PSR J1838-0655, another 'MeV' pulsar}

An interesting discovery turned out to be the detection of soft
$\gamma$-ray emission up to $\sim$300 keV from the ASCA source AX
J1838.0-0655 using {\inte} IBIS/ISGRI data \citep{malizia2005}. Its
location made an association with the TeV source HESS J1837-069
\citep{aharonian2005} plausible, suggesting a pulsar/PWN origin, later
confirmed using RXTE PCA data (23.4 kyr-old pulsar with period 70.5
ms; \citealt{gotthelf2008, kui08}). This young pulsar turned out to be
the third pulsar in hard-X-ray flux after the Crab pulsar and PSR
B1509-58, without a detection in the radio and {\it Fermi} high-energy
$\gamma$-ray bands. Analysis of {\rxte}/HEXTE and {\inte} IBIS/ISGRI
data revealed X-ray pulse profiles up to $\sim$150 keV with a pulse
and spectral shapes similar to that of PSR B1509-58
\citep{kuiper2015}. This pulsar turned out to be another prototype
example of a pulsar with a spectrum reaching maximum luminosity at MeV
energies.

\subsubsection{PSR J1846-0258 showing magnetar-like behaviour}

A very intriguing high-B-field ($4.9 \times 10^{13}$ G) pulsar is PSR
J1846-0258, a relatively slow ({\it P} $\sim$ 324 ms) radio-quiet
pulsar, with the smallest characteristic age ($\tau\sim$ 723 yr) of
all known pulsars. It is located in the centre of SNR Kes75, showing
up as a bright hard X-ray source surrounded by a diffuse PWN. {\inte}
detected point-source emission up to $\sim200$ keV and pulsed emission
up to $\sim 150$ keV (see \citealt{kui09} and references therein). Most
surprisingly, this pulsar showed magnetar-like behaviour during 2006
June 7-12 with an increase by $\sim$ a factor five in luminosity due
to a radiative outburst in soft X-rays lasting 55 days and accompanied
by five magnetar-like bursts \citep{kumar2008, gavriil2008}. The onset
of the radiative event was accompanied by a major spin-up glitch of
the pulsar \citep{kui09}. IBIS/ISGRI could study the evolution of the
total non-thermal flux (PSR J1846-025 + Kes-75) during the years
2003$-$2006 before and after the outburst. The pulsar was found to be
stable in X-rays before the spin-up glitch with a power-law spectrum
(index $\Gamma=$ 1.80 $\pm$ 0.06). During the outburst the hard X-ray
flux increased by $\sim$ 50\%, the spectral shape remained the same,
and after one year the non-thermal emission was back to its
pre-outburst values. The X-ray pulse profile measured by IBIS/ISGRI,
PCA and HEXTE, was a broad single asymmetric pulse that did not vary
in shape over the 3$-$150 keV energy range and, remarkably, did not
change during the magnetar-like outburst, nor did its non-thermal
spectral shape (power-law index $\Gamma \sim$ 1.2). The accurate
IBIS/ISGRI measurement of the total and pulsed spectra showed that the
pulsed fraction approaches 100\% around 150 keV \citep{kui09}. In its
steady state, also this pulsar exhibits very similar timing and
spectral characteristics as PSR B1509-58 with a spectrum reaching its
maximum luminosity at MeV energies, confirmed with the recent
detection with {\it Fermi} LAT of a very weak pulsed emission between
30 and 100 MeV \citep{kuiper2018}.

\subsubsection{Hard X-ray pulsars, a distinct subset of the non-thermal population of rotation-powered pulsars}

The hard X-ray/soft $\gamma$-ray (E$\ge$20 keV) pulsar population
counts only 18 members \citep{kuiper2015}, all non-recycled pulsars,
compared to a total of $\sim$160 (status in 2019) non-recycled pulsars
detected by {\it Fermi} LAT above 100 MeV. As was expected, {\inte}
could not detect the very weak pulsed emission from the recycled MSPs,
even though {\it Fermi} had increased the number of MSPs detected
above 100 MeV from one to $\sim$ 60. The question from the start of
the {\inte} mission was: will {\inte} just detect/confirm the hard
X-ray spectral tails of the high-energy $\gamma$-ray pulsars that
reach their maximum luminosities at GeV energies and have
predominantly narrow (double) pulse profiles, or will it rather
provide new information on the total high-energy non-thermal pulsar
population? In order to investigate this question, the characteristics
of the 18 soft $\gamma$-ray pulsars were compared with those of the 77
non-recycled LAT-detected pulsars in the Second {\it Fermi} Pulsar
Catalog \citep{kuiper2015}.  Surprisingly, it was found that the soft
$\gamma$-ray pulsars are all fast rotators and on average $\sim9.3$
times younger and $\sim43$ times more energetic than the {\it Fermi}
LAT sample (see also \citealt{cze20} for a recent assessment). The
majority (11 sources) exhibits broad, structured single pulse
profiles, and only six have double (or even multiple, Vela) pulses
(see Fig.~\ref{he_psr_morph}). Fifteen soft \gr-ray pulsars show hard
power-law spectra in the hard X-ray band and reach maximum
luminosities typically in the MeV range (see
Fig.~\ref{spectralcompilation}). Pulsed emission has also been
detected by the LAT for only seven of the 18 soft \gr-ray pulsars, but
12 have a PWN detected at TeV energies. In conclusion, observations of
rotation-powered pulsars at hard X-rays and MeV $\gamma$-rays reveal a
subset of the total high-energy pulsar population that can not (yet)
be observed with {\it Fermi} at energies above 100 MeV. This distinct
subsample was originally not taken into account in population studies
based on the {\it Fermi} catalog. However, \citep{torres2018} and
\citep{tor19} recently presented a physical model for the
non-thermal emission of pulsars above 1 keV to fit the spectra of the
$\gamma$/X-ray pulsars along seven orders of magnitude over the
{\inte} band up to the {\it Fermi} high-energy $\gamma$-rays. The
spectra of all pulsars with detected non-thermal emission could be
modeled with a continuous variation of only four model parameters, and
it was proposed that their values likely relate to the closure
mechanism operating in the accelerating region.

After the launch of {\it NuSTAR} in June 2012 (two orders of magnitude more sensitive than {\inte} in the energy band 3$-$79 keV) several of the rotation-powered pulsars in the soft $\gamma$-ray pulsar catalog have been observed. The published temporal and spectral characteristics were all confirmed, and so far no new entries to the pulsar catalog were reported. However, {\it NuSTAR} made some progress by detecting the non-thermal hard-X-ray pulsations of PSR B1821-24, PSR B1937+21 and PSR J0218+4232 for energies up to $\sim$50 keV, $\sim$20 keV and $\sim$25 keV, respectively \citep{gotthelf2017}, and confirming the earlier reported hard X-ray spectra with photon indices $\Gamma\sim$1.1.   Still, {\inte} remains unique in its capabilities to image point sources (pulsars + PWNe) up to few hundred keV, as well as to measure pulsed-emission from non-recycled pulsars above 80 keV.

\section{Magnetars}
\label{sec:magnetars}

Magnetars are isolated NSs defined by the fact that the main   source powering their persistent and flaring emission is   
magnetic energy \citep{tho95,tho96}.  Their external magnetic field is estimated to reach 10$^{15}$ G, and their internal field  might be even higher.  Although we know only about two dozens of magnetars in the Galaxy and in the Magellanic Clouds, the extreme properties of this small class of objects make them particularly interesting as laboratories to study physical processes in high magnetic fields. They emit predominantly in the X-ray and soft $\gamma$-ray energy range, where they show a variety of variable phenomena, ranging from short bursts on sub-second timescales,  to outbursts lasting several months.

Their magnetic field, much larger than in ordinary NSs, is likely produced thanks to a very short rotational period at birth, of the order of only 2$-$3 ms \citep{dun92}.   Magnetars have been invoked as central engines of $\gamma$-ray bursts, able to power   the  prompt emission and/or part of the afterglow,  as a possible explanation for the enigmatic Fast Radio Bursts,  as well as potential sources of gravitational waves.

For a thorough description of the magnetar observational properties and of the theoretical models proposed to explain them we refer to several recent reviews \citep{rea11,MPM15,tur15,kas17}. Here, we concentrate on the results obtained with the \int\  satellite. We first describe the discovery of hard X-ray emission in magnetars,  then we summarize the results concerning the short bursts, and, finally, we describe the observations of the only giant flare emitted by a magnetar after the \int\ launch.

\subsection{The discovery of persistent hard X-ray emission}
\label{sec:HX}

Although magnetars were first discovered as Soft Gamma-ray Repeaters
(SGRs) through the detection of bursts in the hard X-ray band
\citep{maz79}, until the launch of \int\ their persistent emission had
been observed only in the classical $\sim1-10$~keV X-ray range.  At
these energies, the persistent X-ray counterparts of the SGRs, as well
as the Anomalous X-ray Pulsars (AXPs, another class of X-ray sources
later recognized to be magnetars; \citealt{mer95}), are characterized
by rather soft X-ray spectra. These spectra were usually fitted with
the sum of a thermal component, with typical blackbody temperature of
the order of $\sim$0.5 keV, and a steep power law with photon index
$\Gamma$ ranging between $\sim 3$ and $\sim4$ \citep{mer08}.
Therefore, the \int\ discovery of persistent hard X-ray emission from
several magnetars was quite unexpected since a simple extrapolation of
the X-ray spectra would lead to very small hard X-ray fluxes.

An \int\ source with an average flux of 7 mCrab in the 60-120 keV
range, and coincident with the AXP 1E~1841--045 in the Kes 73
supernova remnant, was first reported in \citep{molkov2004}. This
discovery prompted an analysis of archival $RXTE$ data
(\citealt{kui04}, see also \citealt{got07} for a re-analysis of {\it
  BeppoSAX} data) that, thanks to the detections of pulsations up to
$\sim$150 keV, confirmed that the hard X-rays were indeed emitted by
the magnetar and not by the supernova remnant. The pulsations were
later found also in the IBIS/ISGRI data \citep{kui06}.

\begin{figure}[t!]
\centering
\includegraphics[width=\columnwidth]{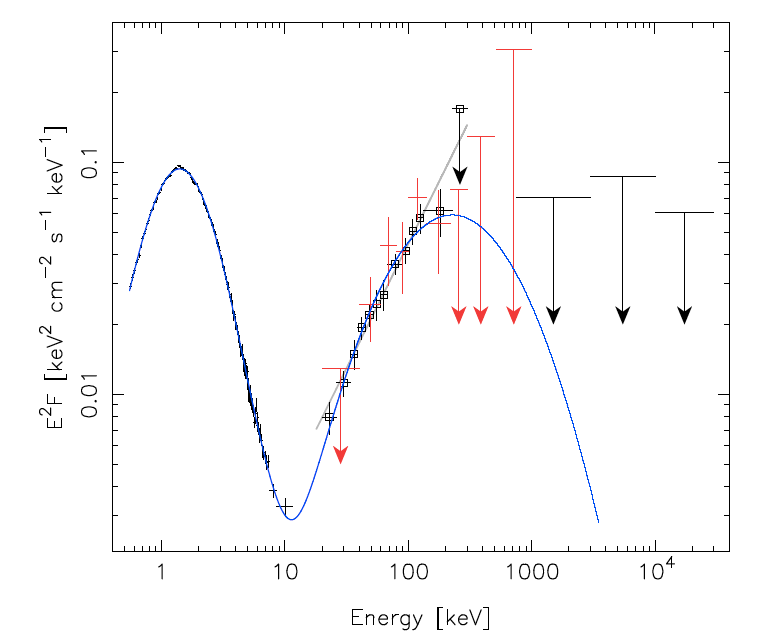}
\caption{The broad band spectrum of 4U~0142+61 (from
  \citep{den08a}). The data below 10 keV are from {\it XMM-Newton},
  while those above 20 keV are from the \int\ IBIS/ISGRI (black) and
  SPI (red) instruments. The upper limits in the 0.75$-$30 MeV range
  are from COMPTEL. The blue line is a best fit with a
  phenomenological model consisting of the sum of log-parabolic
  functions, that clearly illustrates the different components at soft
  and hard X-ray energies. Credit: den Hartog et al., A\&A, 489, 245
  (2008), reproduced with permission (ESO).}
\label{fig:spectrum0142}
\end{figure}

Hard X-ray spectral components were subsequently detected with
\int\ in other AXPs (1RXS~1708$-$4009 and 4U~0142+61;
\citealt{rev04,kui06,den08a}), as well as in the two brightest SGRs:
1806$-$20 \citep{mer05c,mol05,esp07} and 1900+14
\citealt{goe06,duc15}.  As an example, we show in
Fig.~\ref{fig:spectrum0142} the spectrum of 4U~0142+61.  All these
magnetars are persistently bright sources, but in the following years
\int\ detected hard X-ray tails also in transient magnetars, such as
SGR 0501+4516 \citep{rea09} and \qui\ \citep{ber11,kui12}, when they
went in outbursts.

The {\inte} results for the latter source are particularly interesting because  they showed the appearance of a new transient component extending up to 150 keV after the onset of the January 2009 outburst \citep{kui12}. 
This pulsed component was shifted in phase with respect to the lower energy pulse profile and had a different spectral and flux evolution compared to that of the total hard X-ray emission.

\begin{figure}[t!]
\centering
\includegraphics[width=\columnwidth]{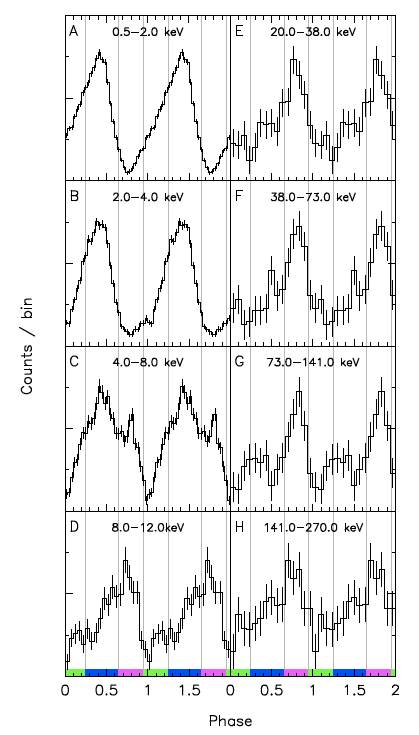}
\caption{Pulse profiles of  1RXS~1708$-$4009  (from \citep{den08b}) obtained in the soft X-ray band  with {\it XMM-Newton} (left panels) and in the hard X-ray band with  {\inte}  (right panels). Observations were not strictly simultaneous. Credit: den Hartog
et al., A\&A, 489, 263 (2008), reproduced with permission (ESO).}
\label{fig:lc1708}
\end{figure}

These \int\ observations, as well as further hard X-ray data  obtained with other satellites  (mainly {\it Suzaku}  and {\it NuSTAR}   \citep{eno10,eno17,an14}), have clearly shown that the magnetar hard X-ray tails extending up to $\sim150-200$~keV are not a simple extrapolation of their lower energy emission. In fact, they are fitted by flatter power-laws ($\Gamma_{\gamma}\sim0.5-2$) and often show a pulse profile different from that seen below 10 keV (see, e.g., Fig.~\ref{fig:lc1708}).  The pulsed flux is generally harder than the unpulsed one, causing an increase of pulsed fraction with energy.  The upper limits in the MeV region \citep{kui06,den08a,den08b} imply a spectral turn-over of the hard components, which, nevertheless, contain an energy of the same order of  that of the soft X-ray emission, or in some cases even larger. 

While the soft X-rays are generally ascribed to thermal emission from the magnetar surface, modified by the effects of a strongly magnetized atmosphere, the hard X-ray tails are thought to originate from non-thermal particles in the magnetosphere (see, e.g., \citep{bel13}), likely with an important contribution from resonant cyclotron scattering \citep{bar07}. In the twisted magnetospheres of magnetars, currents flow along bundles of closed field lines, which in turn can have a complicated geometry with time-dependent local structures. As a result, the computations of the emerging spectra \citep{zan11,bel13b,wad18}, as well as their consistent comparison with the observational data \citep{has14,ten15},  were not simple.  Nevertheless, this research field triggered by the \int\ discovery offered in perspective a new important channel for the understanding of physical properties of magnetars and their emission processes.

\subsection{Bursts from  soft gamma-ray repeaters}
\label{sec:burst}

Two magnetars have been particularly active during the first years of
the \int\ mission, leading to the detection of several bursts:
\sgr\ and \qui . Many of these bursts, as well as a few ones from
other magnetars, occurred within the field of view of the IBIS/ISGRI
instrument and were detected and localized in real time by the
INTEGRAL Burst Alert System (IBAS; \citealt{mer03}).  More recently, a
particularly interesting result was obtained with the IBAS detection
of a peculiar burst from the magnetar SGR 1935$+$2154
\citep{mer20}. This burst was characterized by the simultaneous
emission of an extremely bright radio pulse with properties similar to
those of the fast radio bursts \citep{chi20,sta20}.


 A detailed analysis of the bursts from \sgr\ was reported in
 \citet{goe04} and \citet{goe06b}.  Thanks to the high sensitivity of
 the IBIS/ISGRI imager in the 15-200 keV range, it was possible to
 study for the first time the faint end of the luminosity distribution
 of the SGR bursts. Indeed, the faintest bursts observed in October
 2003 had fluences as low as $2\times10^{-8}$ erg cm$^{-2}$. Several
 bursts showed a significant spectral evolution in the hard X-ray
 range and some evidence for an overall anticorrelation between
 spectral hardness and flux was found with a time resolved spectral
 analysis of the whole sample of bursts \citep{goe04}. This
 anticorrelation was later confirmed with the analysis of a larger
 sample of more than 200 bursts \citep{goe06b}. The {\inte}
 distribution of the burst fluences was found to be a power law with
 index (0.91$\pm$0.09), for fluences in the range between
 $3\times10^{-8}$ and $2\times10^{-6}$ erg cm$^{-2}$.  This rather
 flat slope implies that the integrated flux of fainter bursts below
 the detection threshold does not contribute significantly to the flux
 of the ``persistent'' hard X-ray emission.
 
\begin{figure}[t!]
\centering
\includegraphics[width=\columnwidth]{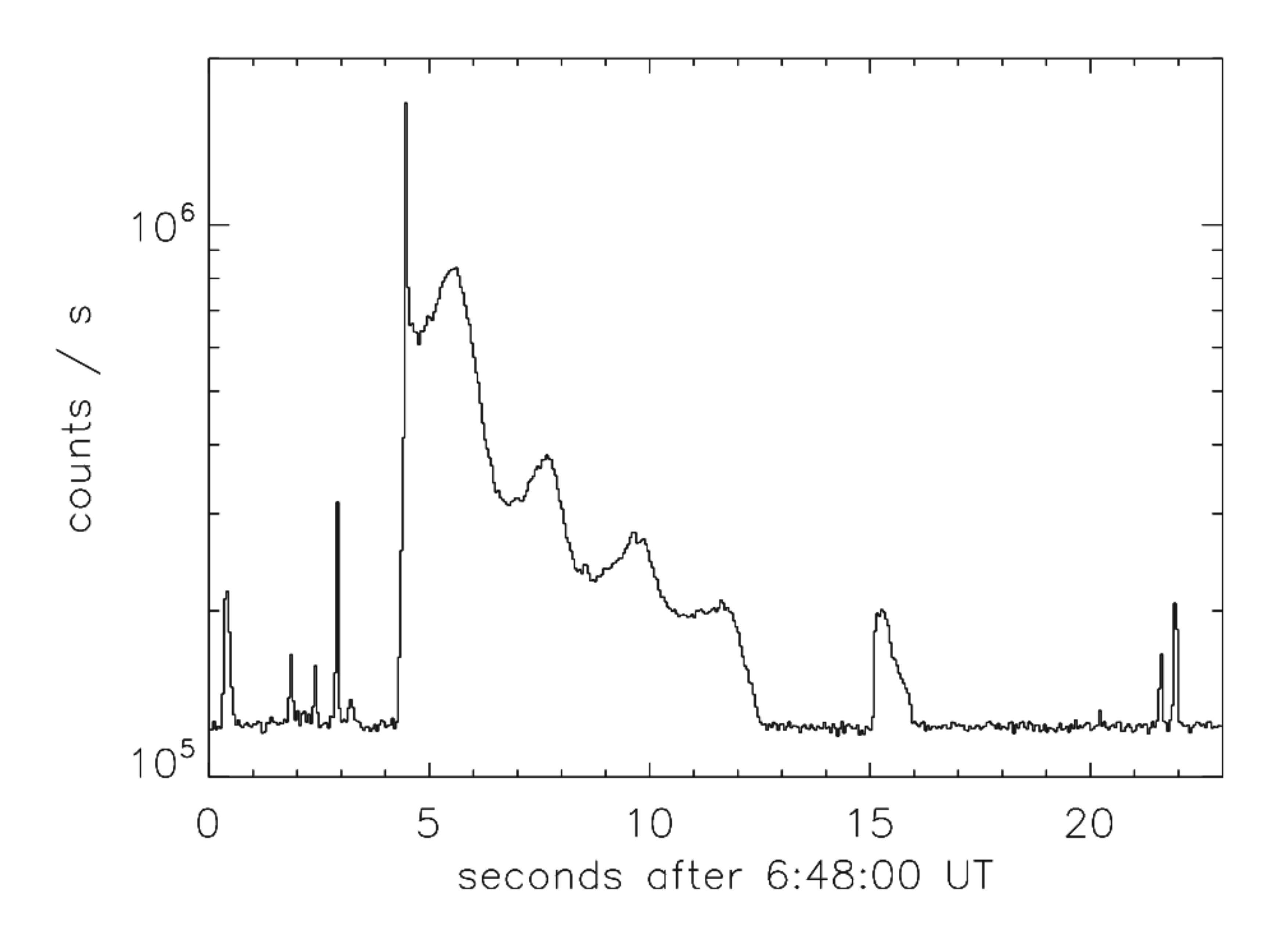}
\caption{The SPI/ACS light curve at $E>80$~keV of a bright burst observed from \qui\ on January 22, 2009. The tail following the bright initial pulse clearly shows the NS rotation period of 2.1 s. Credit: Mereghetti et al.,
ApJ, 696, L74 (2009), reproduced with permission of the AAS.   }
\label{fig:burst1547}
\end{figure}

 \qui\ is a transient magnetar that exhibited three major outbursts: in June 2007, October 2008 and January 2009 \citep{ber11}. During the latter outburst, this source emitted numerous short bursts reaching a particularly  high rate on January 22,
when more than 200 bursts were detected by \int\ in a few hours \citep{mer09}.  Contrary to the case of \sgr , they showed a positive correlation between hardness and intensity \citep{sav10}.
Some of these bursts were particularly bright, reaching a peak flux above
2$\times$10$^{-4}$ erg cm$^{-2}$ s$^{-1}$ at $E>$25 keV. 
Two of them lasted several seconds, and had  tails  modulated at the NS spin period of 2.1 s. 
In particular, the time evolution of the burst shown in Fig.~\ref{fig:burst1547}   resembled that of the giant flares. However,   the energy released in this event was at most $\sim10^{43}$ erg (for $d=5$~kpc), which is orders of magnitude smaller than that of the  three magnetar giant flares observed to date. 
 
The \int\ results on \qui\ triggered follow-up observations with other facilities.
X-ray images obtained on 2009, January 23 and in the following two weeks with {\it Swift}  and {\it  XMM-Newton}  showed the presence of three expanding  rings around the source position \citep{tie10}, caused by   scattering from relatively  thin dust layers along the line of sight.
By fitting the expansion rate of the rings it was possible to determine the   time of the burst(s)  responsible for the scattered X-ray radiation and it was found to be well in agreement with the period of highest bursting activity seen with \int . Furthermore, with a spectral analysis of the scattered X-rays it was also possible to estimate a  distance of 4-5 kpc for  {\qui}.
  
  A particularly interesting burst was discovered at 14:34:24 UTC of
  2020 April 28 by the IBAS software, that automatically identified
  its origin from the transient magnetar SGR 1935$+$2154 and
  distributed a public alert after less than 10 seconds.  This event
  was indepentently discovered at radio wavelengths
  \citep{chi20,sta20} and represents the first, and so far unique, SGR
  burst from which simultaneous radio emission has been detected.  The
  INTEGRAL data obtained with IBIS, as well as those obtained by other
  high-energy satellites \citep{li20,rid20}, showed that its spectrum
  was harder than that of typical magnetar bursts. On the other hand,
  this event was not particularly luminous, with a 20-200 keV emitted
  energy of the order of $\sim 10^{39}$ erg (assuming isotropic
  emission and a distance of 4.4 kpc; \citealt{mer20}).  In the
  400-800 MHz band the burst consisted of two narrow pulses separated
  by 29 ms and with a total fluence of 700 kJy ms. Interestingly, two
  narrow pulses with the same separation, are visible also in the IBIS
  light curve and have a delay of 6.5 ms with respect to the radio
  ones.  The discovery of simultaneous fast bursting emission at radio
  and high-energies from SGR 1935$+$2154 gives strong support to
  models based on magnetars that have been proposed to explain the
  enigmatic class of sources known as fast radio bursts \citep{pet19,
    pla19}.

\subsection{The 2004 giant flare from SGR 1806$-$20}
\label{sec:GF}

On  December 27, 2004,  a very bright burst was detected in the Anti-Coincidence Shield (ACS) of the SPI instrument  and  the corresponding light curve derived by the   IBAS software was automatically published on-line in real time.  Due to the large peak flux reached in the first 200 ms of the burst, the following tail was almost invisible on a linear scale, but, after a closer  examination, its presence was noticed. The clear periodicity at 7.6 s visible in the burst tail unequivocally identified this event as a giant flare from \sgr\ \citep{bor04},  similar to the  two giant flares observed on 1979 March 5 from SGR 0525$-$66
and  on 1998 August 27 from SGR 1900+14. 
This  giant flare,  first reported by \int , was the culmination of  two years of increasing bursting activity  from \sgr , that was accompanied by a spectral hardening and an increase in the spin-down rate \citep{mer05a,goe06b}.

Despite the downward revision in the source distance (from an initial
estimate of 15 kpc, to the most likely value of 8.7 kpc;
\citealt{bib08}) the \sgr\ giant flare has been the most energetic
observed so far, with a peak isotropic luminosity of $\sim10^{47}$ erg
s$^{-1}$.  The total energy release of $\sim10^{46}$ erg implies a
catastrophic magnetic reconnection, associated to a major crustal
fracture, and leading to a global reconfiguration of the NS's magnetic
field.

\begin{figure}[t!]
\centering
\includegraphics[width=\columnwidth]{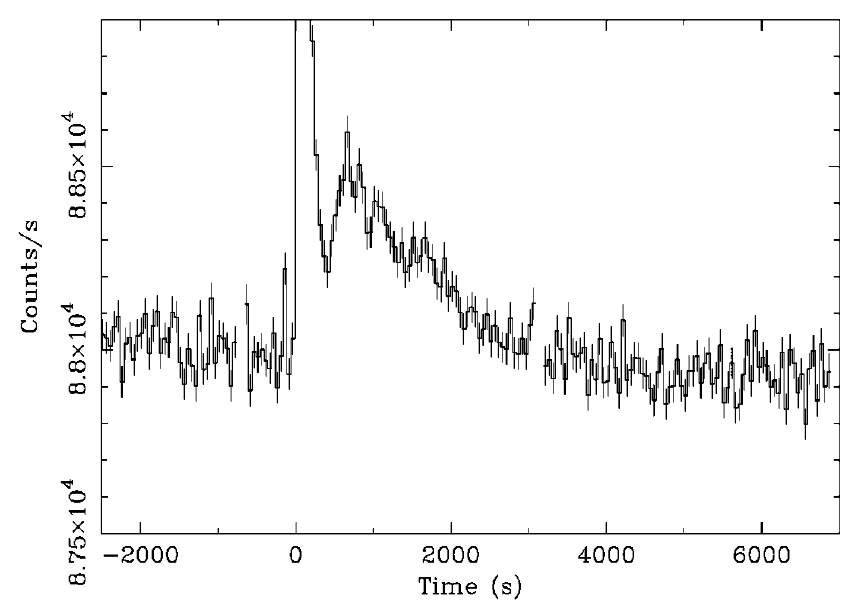}
\caption{SPI/ACS light curve ($E>80$~keV) of the December 27, 2004
  giant flare from \sgr\ (from \citealt{mer05b}).  Due to the
  rebinning at 50 s, the 7.6 s pulsations in the tail time interval
  (1$-$400 s) are not visible in this figure. After the pulsating
  tail, the flux increases again, peaking $\sim$700 s after the
  beginning of the flare and decreasing below the background level
  about one hour later.  Note that the peak of the flare at $t=0$,
  reaching an observed count rate $>2\times10^{6}$ counts s$^{-1}$, is
  out of the vertical scale.  Credit: Mereghetti et al., ApJ, 624,
  L105 (2005), reproduced with permission of the AAS.  }
\label{fig:1806lc}
\end{figure}

Detailed results on the 2004 giant flare were obtained with different satellites \citep{pal05,ter05,hur05}. In addition to the features reported in these works, thanks to the   very large effective area of the ACS,   \int\ could discover a long-lasting    emission that might be the first evidence for a soft $\gamma$-ray afterglow following a SGR giant flare \citep{mer05b}. 
As it can be seen in Fig.~\ref{fig:1806lc}, after the end of the pulsating tail, the ACS count rate increased again, reaching a peak at $t\sim700$ s, and then returned to the pre-flare background level at $t\sim3000-4000$~s. 
The presence of this long-lasting emission was later confirmed with 
{\it Konus-Wind} \citep{fre07} and {\it RHESSI} \citep{bog07} data, although with smaller statistics and on different time intervals.

The time evolution of the ``afterglow'' component seen by the \int\ ACS above $\sim$80 keV is well fitted by a power-law decay with $F(t)\propto  t^{-0.85}$. 
For  a thermal Bremsstrahlung spectrum with
temperature $kT=30$~keV, the fluence at  $E>80$~keV in the
400-4000 s time interval was $\sim3\times10^{-4}$ erg cm$^{-2}$, which is of the same order as that in the pulsating tail (i.e. in the 1-400 s time interval).
However, since the ACS does not provide spectral information, an estimate of the total energy, i.e. including the contribution at energies below $\sim$80 keV,  is affected by a significant   uncertainty due to the spectral extrapolation.
    
The power-law time evolution, as well as the hard power-law spectrum
(photon index $\Gamma\sim$1.6; \citealt{fre07}), suggest to interpret
this long-lasting emission as radiation caused by the interaction of
relativistic ejecta from \sgr\ with the circumstellar material
\citep{mer05b}, similar to the afterglows seen in $\gamma$-ray bursts.
With standard models for $\gamma$-ray burst afterglows based on
synchrotron emission, it is possible to relate the bulk Lorentz factor
of the ejected material, $\gamma_{ej}$, with the time $t_{0}$ of the
afterglow onset. The values observed with \int\ give
$\gamma_{ej}\sim$15($E$/5$\times$ 10$^{43}$ erg)$^{1/8}$($n$/0.1
cm$^{-3}$)$^{-1/8}$($t_{0}$/100\,\mbox{s})$^{-3/8}$, where $n$ is the
ambient density. $\gamma_{ej}$ is thus consistent with a mildly
relativistic outflow, as also inferred from the analysis of the radio
source that appeared after the giant flare \citep{gra06}.

\section{Future prospectives}

{\inte} has been successfully operated in Space for almost 18 years, and no major degradation of the instrument capabilities has been recorded so far. In principle, the mission scientific operations could continue until 2029, when the satellite is already planned to re-enter the Earth atmosphere. As emphasized multiple times in this review, the unique combination of sensitivity, timing resolution, large field of view and angular resolution of the {\inte} instruments has led to crucial advancements in our understanding of the pulsating hard X-ray sky.  

For the study of the AMSPs and transitional millisecond X-ray pulsars in outburst, {\inte} will certainly continue to provide the means to discover additional rare members of these classes of sources.  Any newly discovered object has provided unique insights into the accretion physics, leading to advancements in the understanding of the interaction between the accretion flow and the intense magnetic/gravitational field of the NS, as well as the enrichment of the interstellar medium through the thermonuclear explosions that often go off close to the compact object surface. So far, only one transitional X-ray pulsar has been caught during a  bright X-ray outburst state, indeed thanks to  {\inte} observations.  It is thus of paramount importance to continue hunting for these peculiar objects in order to solve the unknown mechanisms driving their atypical behavior. Similar considerations hold for other transient NS sources, as the magnetars. The past 18 years of observations have proven that these objects are well at reach for the {\inte} instrumentation and we expect that additional outbursting events revealed during the {\inte} monitoring of the sky will produce rich data-sets to improve our understanding of the complex magnetospheric phenomenology that is driving the high energy emission from these systems. 

The complementarity of the instruments on-board {\inte} with those of other last generation X-ray facilities, as {\xmm}, {\chandra}, {\nustar}, and {\nicer}, proved fundamental to go beyond the simple identification of new isolated and/or binary systems with rapidly rotating NSs. It also allowed to dig deeply inside the properties of their high energy emission, with {\inte} providing a unique contribution in the hardest X-ray energy band ($\gtrsim$80~keV). Efforts are on-going to improve the rapidity of the response of the different facilities to the frequent {\inte} discoveries and to coordinate sub-sequent observations in a multi-messenger fashion \citep[see, e.g.,][and references therein]{middleton}. 

Continued observations of the hard X-ray sky with {\inte} in the coming years will also certainly improve our sensitivity to detect fainter and fainter new hard sources. This will clearly increase the chances of discovering new multi-wavelength sources, such as transitional millisecond pulsars. In this case, the {\inte} contribution  will help us address fundamental questions such as which evolutionary channels produce transitional objects, how frequent is the occurrence of the faint disk state observed from these systems and ultimately which physical mechanism powers it and makes these objects different from standard AMSPs. Based on this and similar past experiences, we expect that the improved synergies between {\inte} and the other operating facilities in the multi-messenger context will help unveil the true nature of the newly discovered  sources, providing further exciting discoveries and unexpected challenges in the decade to come.

\section{Acknowledgements}

Al.~P., D.~d.~M. and T.~D.~S. acknowledge financial support from ASI/INAF I/037/12/0, ASI/INAF 2017-14-H.0 (PI: De Rosa, PI: Belloni) and from INAF 
"Sostegno alla ricerca scientifica main streams dell'INAF", Presidential Decree 43/2018 and from  "SKA/CTA projects", Presidential Decree N. 70/2016.
J.P. was supported by the Ministry of Science and Higher Education of the Russian Federation grant 14.W03.31.0021. The INTEGRAL French teams acknowledge partial funding from the French Space Agency (CNES). Z.L. thanks the International Space Science Institute in Bern for the hospitality. Z.L was supported by National Natural Science Foundation of China  (Grant Nos. U1938107, 11703021, U1938107, 11873041). Ad.~P. and S.~M. acknowledge continuous support from the Italian Space Agency ASI through the ASI/INAF agreement n.2019-35-HH. D.~F.~T and the group at the Institute of Space Sciences acknowledges support via grants PGC2018-095512-B-I00, SGR2017-1383, andAYA2017-92402-EXP. F~.C.~Z. is supported by a Juan de la Cierva fellowship. V.~D.~F. acknowledges Silesian University in Opava and Gruppo Nazionale di Fisica Matematica of Istituto Nazionale di Alta Matematica for support. This research was supported by the Polish National Science Centre grant number 2017/25/B/ST9/02805. J.~L. acknowledges the support
from the Alexander von Humboldt Foundation.

\section*{References}
\bibliographystyle{elsarticle-harv}
\bibliography{mybibfile}

\end{document}